\documentclass[11pt,a4paper]{article}
\pdfoutput=1
\usepackage{jheppub}
\usepackage[utf8]{inputenc}
\usepackage[english]{babel}
\usepackage{booktabs}

\title{Precise prediction for the W boson mass in the MRSSM}

\author[a,1,2]{Philip Diessner,\note{Former address.}\note{Corresponding author. 
}}
\author[a]{Georg Weiglein}
\affiliation[a]{Deutsches Elektronen-Synchrotron DESY, Hamburg, Germany}
\emailAdd{philip.diessner@desy.de}
\emailAdd{georg.weiglein@desy.de}
\preprint{DESY 19-035}
\abstract{The mass of the W boson, $\MW$, plays a central role for
high-precision tests of the electroweak theory. Confronting precise
theoretical
predictions with the accurately measured experimental value provides a high
sensitivity to quantum effects of the theory entering via loop
contributions. The currently most accurate prediction for the W boson mass
in the Minimal R-symmetric Supersymmetric Standard Model (MRSSM) is
presented. Employing the on-shell scheme, it combines all numerically
relevant contributions that are
known in the Standard Model (SM) in a consistent way with all MRSSM one-loop
corrections. Special care is taken in the treatment of the triplet scalar
vacuum expectation value $v_T$ that enters the prediction for $\MW$ already
at lowest order. In the numerical analysis the decoupling properties of the
supersymmetric loop contributions and the comparison with the MSSM are
investigated. Potentially large numerical effects of the MRSSM-specific
$\Lambda$ superpotential couplings are highlighted. The comparison with
existing results in the literature is discussed.
 }

\newcommand{\MW}{M_W}
\newcommand{\MZ}{M_Z}
\newcommand{\lsim}
{\;\raisebox{-.3em}{$\stackrel{\displaystyle <}{\sim}$}\;}
\newcommand{\gsim}
{\;\raisebox{-.3em}{$\stackrel{\displaystyle >}{\sim}$}\;}

\newcommand{\cmass}{\ensuremath{c_W^2}}
\newcommand{\cmix}{\ensuremath{\tilde c_W}}
\newcommand{\cmixx}{\ensuremath{\tilde c_W^2}}
\newcommand{\smas}{\ensuremath{s_W}}
\newcommand{\smass}{\ensuremath{s_W^2}}
\newcommand{\smix}{\ensuremath{\tilde s_W}}
\newcommand{\smixx}{\ensuremath{\tilde s_W^2}}
\newcommand{\Dr}{\ensuremath{\Delta r}}
\newcommand{\Dhr}{\ensuremath{\Delta \tilde r}}
\newcommand{\MS}{\ensuremath{\overline{\text{MS}}}}
\newcommand{\DR}{\ensuremath{\overline{\text{DR}}}}
\newcommand{\muu}[1]{\mu_u^{\text{eff,}#1}}
\newcommand{\mud}[1]{\mu_d^{\text{eff,}#1}}
\DeclareMathOperator{\Tr}{Tr}
\begin{document}
\maketitle
\flushbottom
\section{Introduction}
Supersymmetry (SUSY) is seen as one of the most attractive
extensions of the Standard Model (SM) of particle physics. 
It provides a solution for the hierarchy problem of the SM and a prediction
for the mass of the scalar resonance  discovered at the 
LHC~\cite{ATLAS:2012gk,Chatrchyan:2012ufa,Aad:2015zhl}
if it is appropriately identified with one of the neutral Higgs bosons of
the considered supersymmetric model.

So far, no further new state has been discovered at the LHC. The search
limits from the LHC and previous colliders give rise to 
strong constraints on the parameter space
of the Minimal Supersymmetric Standard Model (MSSM). The impact of those
search limits may be much smaller for models with extended SUSY sectors. In
particular, the introduction of R-symmetry~\cite{Kribs:2012gx} leads to
models where the limits on the particle masses are significantly weaker and
where accordingly the 
discovery of TeV-scale SUSY is still in reach of current experiments.

The LHC phenomenology of the Minimal R-symmetric Supersymmetric Standard Model (MRSSM)~\cite{Kribs:2007ac} and other models with R-symmetry has 
recently been explored in refs.~\cite{PD1,PD2,PD3,PD4,Kotlarski:2016zhv,
Beauchesne:2017jou,Benakli:2018vqz,Alvarado:2018rfl,Darme:2018dvz,Chalons:2018gez}.
The signatures of the MRSSM differ from the ones of the MSSM
as the model includes Dirac gauginos and higgsinos instead of Majorana ones.
This goes hand in hand with additional states in the Higgs sector
including an SU$(2)$ triplet, which may acquire a vacuum expectation value
breaking the custodial symmetry of the electroweak sector.
Additionally, R-symmetry forbids all soft SUSY-breaking trilinear couplings 
between Higgs bosons and squarks or sleptons, which removes unwanted sources 
of flavour violation which exist in the MSSM. 

Even with no direct signals of SUSY indirect probes like electroweak precision 
observables can provide sensitivity to SUSY contributions above the direct 
experimental reach.
Here we study the prediction for the mass
of the W gauge-boson, $M_W$. 
The world average combining LEP and Tevatron results~\cite{Schael:2013ita,Aaltonen:2013iut} is
\begin{equation}
M_W^{\text{exp.}}=80.385\pm0.015 \text{ GeV}\;.
\label{eq:mw_value_exp}
\end{equation}
It may be improved by measurements of the LHC experiments, where a first
ATLAS result at $\sqrt{s}=7$~TeV reports a value~\cite{Aaboud:2017svj} of
\begin{equation}
M_W^{\text{ATLAS}}=80.370\pm0.019 \text{ GeV}\;.
\label{eq:mw_value_atlas}
\end{equation}

The statistical combination of the ATLAS measurement~\eqref{eq:mw_value_atlas} with the one of eq.~\eqref{eq:mw_value_exp}
yields an updated experimental result of~\cite{PDG:2018}
\begin{equation}
M_W^{\text{LEP+Tevatron+ATLAS}}=80.379\pm0.012 \text{ GeV}\;.
\label{eq:mw_value_exp_all}
\end{equation}

For a meaningful comparison to experiment a precise theory prediction is necessary.
In the Standard Model (SM) the prediction includes contributions at the
one-loop~\cite{Sirlin:1980nh,Marciano:1980pb} and two-loop level~\cite{Djouadi:1987gn,Djouadi:1987di,Kniehl:1989yc,Halzen:1990je,
Kniehl:1991gu,Kniehl:1992dx,Freitas:2000gg,Freitas:2002ja,Awramik:2002wn,
Awramik:2003ee,Onishchenko:2002ve,Awramik:2002vu}, as well as leading three-
and four-loop corrections~\cite{Avdeev:1994db,Chetyrkin:1995ix,Chetyrkin:1995js,Chetyrkin:1996cf,
Faisst:2003px,vanderBij:2000cg,Boughezal:2004ef,Boughezal:2006xk,Chetyrkin:2006bj,
Schroder:2005db}. 
In refs.~\cite{Awramik:2003rn,Awramik:2006uz} a
parametrisation of the $M_W$ prediction containing all known higher-order 
corrections in the on-shell scheme has been given. Updating the experimental
results for the input values~\cite{PDG:2018}\footnote{See 
section~\ref{sec:results} for the numerical values.} and identifying the scalar 
state discovered at the LHC with the Standard Model Higgs boson leads to a 
prediction for the W boson mass in the Standard Model of
\begin{equation}
M_W^{\text{SM,on-shell}}=80.356\text{ GeV}\,.
\label{eq:mw_value_sm}
\end{equation}
This result shows a slight downward shift compared to the previous 
calculation~\cite{Stal:2015zca} due to the updated input parameters. 
Hence, the long-standing tension
of the experimental measurement and theoretical prediction of just below
$2\sigma$ remains. 

As similar result for the W boson mass prediction in the SM has been
obtained for a 
mixed on-shell/$\overline{\text{DR}}$
scheme in ref.~\cite{Degrassi:1990tu} and updated in ref.~\citep{Degrassi:2014sxa}.

The largest parametric uncertainty is induced by the top quark mass. 
For the experimental value
\begin{equation}
m_t^{\text{exp.}}=173.0\pm0.4 \text{ GeV}\;
\label{eq:mtexp}
\end{equation}
the quoted experimental error accounts only for the uncertainty in
extracting the measured parameter. This needs to be supplemented with the
systematic uncertainty arising from 
relating the measured mass parameter to a theoretically well-defined
quantity such as the $\overline{\text{MS}}$ mass. This uncertainty could
be reduced very significantly with a measurement of the top quark mass from
the $t \bar t$ threshold at a future $e^+e^-$ collider

Accurate theoretical predictions for the W boson mass have also been
obtained  for the most popular supersymmetric 
extensions of the SM, in particular for the 
MSSM~\cite{Heinemeyer:2006px,Heinemeyer:2013dia,Stal:2015zca},
see also ref.~\cite{Heinemeyer:2004gx} and references therein,
and the NMSSM~\cite{Stal:2015zca},  in the on-shell scheme. Furthermore,
results are available in the mixed 
on-shell/$\overline{\text{DR}}$ scheme for the MSSM first studied in
ref.~\cite{BPMZ}, which has been adapted for mass spectrum
generators e.g. \texttt{SPheno}~\cite{Porod:2003um} and \texttt{SoftSUSY}~\cite{Allanach:2001kg} which also includes extensions to the NMSSM~\cite{Allanach:2013kza}. 

For the MRSSM the W boson mass has been studied together 
with other electroweak
precision observables in ref.~\cite{PD1}. The MRSSM result of 
ref.~\cite{PD1} was obtained for the aforementioned
mixed on-shell/$\overline{\text{DR}}$
scheme
where only the gauge-boson masses $\MW$ and $\MZ$ are on-shell quantities,
making use and adapting
the tools \texttt{SARAH} and \texttt{SPheno}~\cite{Staub:2008uz,Staub:2009bi,
Staub:2010jh,Porod:2011nf,Staub:2012pb,Staub:2013tta,Goodsell:2014bna,
Goodsell:2015ira}.
Recently, a new implementation of this calculation in the program
\texttt{FlexibleSUSY}~2.0~\cite{Athron:2017fvs} has shown a large
discrepancy in the MSSM with the result that was achieved in the on-shell
scheme~\cite{Heinemeyer:2006px,Heinemeyer:2013dia,Stal:2015zca} and also a
large discrepancy in the MRSSM with the result of 
ref.~\cite{PD1}. 

In this paper we present an improved prediction for the W boson mass in the
MRSSM. Employing the on-shell scheme for the SM-type parameters, 
we obtain the complete one-loop
contributions in the MRSSM and combine them with the state-of-the-art SM-type
corrections up to the four-loop level. In the calculation of the MRSSM
contributions a renormalisation of the triplet scalar vacuum expectation 
value $v_T$ is needed since this parameter enters the prediction for $\MW$ 
already at lowest order. We investigate the treatment of this parameter and
implement a $\overline{\text{DR}}$-type renormalisation. In our numerical
analysis we study the decoupling limit where
the SUSY particles are heavy and verify that the SM prediction is
recovered in this limit. We investigate the possible size of the different MRSSM
contributions and compare the results with the MSSM case. We also discuss
the comparison of our result with the existing MRSSM results. 

The paper is organised as follows: In section~\ref{sec:MRSSM} we give
a brief overview of the MRSSM field content and introduce the relevant model
parameters required to discuss the calculation of the $M_W$ prediction in
this model.
In section~\ref{sec:dr} details on the calculation of the higher-order
corrections to the muon decay process are presented. 
Section~\ref{sec:implementation} contains the details on the implementation
of the calculation, while in section~\ref{sec:results} we present a 
quantitative study of the $M_W$ prediction in the MRSSM parameter space 
and a comparison of our results to previous calculations. We conclude
in section~\ref{sec:conclusions}.

\section{The Minimal R-symmetric Supersymmetric Standard Model}
\label{sec:MRSSM}
\subsection{Model overview}
The minimal R-symmetric extension of the MSSM, the MRSSM,  requires the introduction of Dirac 
mass terms for gauginos and higgsinos, since Majorana mass terms are forbidden by R-symmetry. 
This leads to the need for an extended number of chiral superfields containing the necessary 
additional fermionic degrees of freedom.

Therefore, the field content of the  MRSSM  is enlarged compared to the MSSM by
doublet superfields ${\hat{R}}_{d,u}$ carrying R-charge under the R-symmetry 
as well as adjoint chiral 
superfields  $\hat{\cal O}$, $\hat{T}$, $\hat{S}$ for each of the gauge groups.
The full field content of the MRSSM including the assignment of R-charges is given in table~\ref{tab:Rcharges}.
%%%%%%%%%%%%%%%%%%%%%%%%%%%%%%%%%%%%%%%%%%%%%%%%%%%%%%%%%%%%%%%%%%%%%%%%%%%%%
\begin{table}[th]
\begin{center}
\begin{tabular}{cllllll}
%\hline
\toprule
\multicolumn{1}{c}{Field} & \multicolumn{2}{c}{Superfield} &
                              \multicolumn{2}{c}{Boson} &
                              \multicolumn{2}{c}{Fermion} \\
\midrule
 \phantom{\rule{0cm}{5mm}}Gauge    &\, $\hat{g},\hat{W},\hat{B}$        \,& \, $\;\,$ 0 \,
          &\, $g,W,B$                 \,& \, $\;\,$ 0 \,
          &\, $\tilde{g},\tilde{W}\tilde{B}$             \,& \, +1 \,  \\
Matter   &\, $\hat{l}, \hat{e}$                    \,& \,\;+1 \,
          &\, $\tilde{l},\tilde{e}^*_R$                 \,& \, +1 \,
          &\, $l,e^*_R$                                 \,& $\;\;\,$\,\;0 \,    \\
          &\, $\hat{q},{\hat{d}},{\hat{u}}$       \,& \,\;+1 \,
          &\, $\tilde{q}_L,{\tilde{d}}^*_R,{\tilde{u}}^*_R$ \,& \, +1 \,
          &\, $q_L,d^*_R,u^*_R$                             \,& $\;\;\,$\,\;0 \,    \\
 H-Higgs    &\, ${\hat{H}}_{d,u}$   \,& $\;\;\,$\, 0 \,
          &\, $H_{d,u}$               \,& $\;\;\,$\, 0 \,
          &\, ${\tilde{H}}_{d,u}$     \,& \, $-$1 \, \\ \hline
\phantom{\rule{0cm}{5mm}} R-Higgs    &\, ${\hat{R}}_{d,u}$   \,& \, +2 \,
          &\, $R_{d,u}$               \,& \, +2 \,
          &\, ${\tilde{R}}_{d,u}$     \,& \, +1 \, \\
  Adjoint Chiral  &\, $\hat{\cal O},\hat{T},\hat{S}$     \,& \, $\;\,$ 0 \,
          &\, $O,T,S$                \,& \, $\;\,$ 0 \,
          &\, $\tilde{O},\tilde{T},\tilde{S}$          \,& \, $-$1 \,  \\
\bottomrule
\end{tabular}
\end{center}
\caption{The R-charges of the superfields and the corresponding bosonic and
             fermionic components.
        }
\label{tab:Rcharges}
\end{table}
%%%%%%%%%%%%%%%%%%%%%%%%%%%%%%%%%%%%%%%%%%%%%%%%%%%%%%%%%%%%%%%%%%%%%%%%%%%%%%

In the following we introduce the parameters of the MRSSM. All model parameters  are taken to be real for the purpose of this work.
The MRSSM superpotential is given as
\begin{align}
\nonumber \mathcal{W} = & \mu_d\,\hat{R}_d \cdot \hat{H}_d\,+\mu_u\,\hat{R}_u\cdot\hat{H}_u\,+\Lambda_d\,\hat{R}_d\cdot \hat{T}\,\hat{H}_d\,+\Lambda_u\,\hat{R}_u\cdot\hat{T}\,\hat{H}_u\,\\ 
 & +\lambda_d\,\hat{S}\,\hat{R}_d\cdot\hat{H}_d\,+\lambda_u\,\hat{S}\,\hat{R}_u\cdot\hat{H}_u\,
 - Y_d \,\hat{d}\,\hat{q}\cdot\hat{H}_d\,- Y_e \,\hat{e}\,\hat{l}\cdot\hat{H}_d\, +Y_u\,\hat{u}\,\hat{q}\cdot\hat{H}_u\, ,
\label{eq:superpot}
 \end{align} 
where the dot denotes the $\epsilon$ contraction with $\epsilon_{12} = +1$.  
In order to achieve canonical kinetic terms the triplet is defined as
\begin{equation}
\hat{T}=
\begin{pmatrix}
\hat{T}_0/\sqrt{2}&\hat{T}_+\\
\hat{T}_-&-\hat{T}_0/\sqrt{2}\\
\end{pmatrix}
\,.
\label{eq:t-repr}
\end{equation}
The usual $\mu$ term of the MSSM is forbidden by R-symmetry but
similar terms can be written down connecting the Higgs doublets to the
inert R-Higgs fields with the parameters $\mu_{d,u}$. Trilinear couplings $\Lambda_{d,u}$ and $\lambda_{d,u}$ couple the doublets to the adjoint triplet and
singlet field, respectively. The Yukawa couplings $Y_{d,e,u}$ are the same as in the MSSM.

The soft SUSY-breaking Lagrangian contains the soft masses for all scalars 
as well as the usual $B_\mu$ term. Trilinear A-terms of the MSSM are forbidden by R-symmetry.\footnote{
Following the arguments in ref.~\cite{PD1}, the addition of further bilinear and trilinear holomorphic terms of the adjoint scalar
fields is not considered here.} This part of the Lagrangian reads
\begin{equation}
  \begin{split}
  \mathcal{L}^{\text{MRSSM}}_{\text{soft}} =
    &- m^2_{\tilde{q}_{L},ij} \tilde{q}_{iL}^*\tilde{q}_{jL}
    - m^2_{\tilde{u}_{R},ij} \tilde{u}_{iR}^*\tilde{u}_{jR}
    - m^2_{\tilde{d}_{R},ij} \tilde{d}_{iR}^*\tilde{d}_{jR}
    - m^2_{\tilde{\ell}_{L},ij} \tilde{\ell}^*_{iL}\tilde{\ell}_{jL}
    - m^2_{\tilde{e}_{R},ij} \tilde{e}^*_{iR}\tilde{e}_{jR}
    \\
    &- m^2_{H_{d}} |H_{d}|^2
    - m^2_{H_{u}} |H_{u}|^2    
    - \left[ B_\mu (H_d H_u) + \text{h.\,c.}\right]
    - m^2_{S} |S|^2\\
    &- m^2_{R_{d}} |R_{d}|^2
    - m^2_{R_{u}} |R_{u}|^2
    - m^2_{T} \Tr(T^* T)
    - m^2_{O} \Tr(O^* O)     
    \;.
  \end{split}
  \label{eq:softmasses}
\end{equation}
Dirac mass terms connecting the gauginos and the fermionic components of 
the adjoint superfields are introduced.
They are generated from the R-symmetric operator involving the D-type spurion \cite{Fox:2002bu} \footnote{Alternatively, it is also possible to generate 
Dirac gaugino masses via F-term breaking~\cite{Martin:2015eca}.
}
\begin{equation}
\label{eq:mdirac}
\int d^2\theta\frac{\hat{W'}_{\alpha,i}}{M} W_i^\alpha \hat{\Phi}_i  
\ni M_i^D\, \tilde{g}_i \tilde{g}'_i,
\end{equation}
where $M$ is the mediation scale of SUSY breaking , $W_i^\alpha$ represents 
the gauge superfield strength tensors, $\tilde{g}_i=\tilde{g},\tilde{W},
\tilde{B}$ is the gaugino, and $\tilde{g}'_i=\tilde{O},\tilde{T},\tilde{S}$  
is the corresponding Dirac partner with opposite R-charge which is part of
a chiral superfield $\hat{\Phi}_i=\hat{\cal O}$, $\hat{T}$, $\hat{S}$.  
The mass term is generated by the spurion field strength $\hat{W'}_{\alpha,i}=\theta_\alpha \mathcal{D}_i$
getting a vev $\langle \mathcal{D}_i\rangle=M M_i^D$.
Additionally, integrating out the spurion in eq.~\eqref{eq:mdirac} generates 
terms coupling the D-fields to the scalar components of the chiral superfields, 
which leads to the appearance of Dirac masses also in the Higgs sector,
\begin{align}
V_D= \, & M_B^D (\tilde{B}\,\tilde{S}-\sqrt{2} \mathcal{D}_B\, S)+
M_W^D(\tilde{W}^a\tilde{T}^a-\sqrt{2}\mathcal{D}_W^a T^a)+
M_g^D(\tilde{g}^a\tilde{O}^a-\sqrt{2}\mathcal{D}_g^a O^a)
 %-\sqrt{2} (M_B^D \mathcal{D}_B S + M_W^D \mathcal{D}_W^a T^a 
+ \mbox{h.c.}\,.
\label{eq:potdirac}
\end{align}

During electroweak symmetry breaking (EWSB) the neutral EW scalars with no R-charge develop vacuum expectation values (vevs)
\begin{align*} 
H_d^0=& \, \frac{1}{\sqrt{2}} (v_d + \phi_{d}+i  \sigma_{d}) \;,& 
H_u^0=& \, \frac{1}{\sqrt{2}} (v_u + \phi_{u}+i  \sigma_{u}) \;,\\ 
T^0  =& \, \frac{1}{\sqrt{2}} (v_T + \phi_T +i  \sigma_T)   \;,&
S   = & \, \frac{1}{\sqrt{2}} (v_S + \phi_S +i  \sigma_S)   \;. 
\end{align*} 
After EWSB the singlet and triplet vevs
effectively shift the $\mu$-parameters of the model, and it is useful
to define the abbreviations
\begin{align}
\mu_i^{\text{eff,}\pm}&
=\mu_i+\frac{\lambda_iv_S}{\sqrt2}
\pm\frac{\Lambda_iv_T}{2}
,
&\mu_i^{\text{eff,}0}&
=\mu_i+\frac{\lambda_iv_S}{\sqrt2}
,&i=u,d.
\end{align}

The $\text{R}=0$ Higgs sector contains four CP-even and three CP-odd neutral as 
well as three charged Higgs bosons.
The Higgs doublets with R-charge 2 stay inert and do not receive a vev such 
that two additional complex neutral and charged scalars are predicted by the MRSSM.
The sfermion sector is the same as in the MSSM with the
restriction that mixing between the left- and right-handed sfermions 
is forbidden by R-symmetry.
In the MRSSM the number of chargino and higgsino degrees of freedom is doubled compared to the MSSM
as the neutralinos are Dirac-type and there are two separated chargino sectors
where the product of electric and R-charge is either 1 or $-1$.

The mass matrices of the SUSY states can be found in ref.~\cite{PD1}.
The masses of the gauge bosons arise as usual with an important distinction
which will be discussed in more detail in the following.

\subsection{The W boson mass in the MRSSM}
The expression for the W boson mass differs from the usual form of the MSSM 
and the SM due to the triplet vev $v_T$.
With the representation in eq.~\eqref{eq:t-repr} the kinetic term for the scalar triplet is given as
\begin{equation}
\mathcal{L}_{T,\text{kin.}} = \text{Tr}\left[ (\mathcal{D}_\mu T)^\dagger(\mathcal{D}^\mu T)\right]\,,
\mathcal{D}_\mu T = \partial_\mu T + \imath g_2 \left[W,T\right]\;.
\end{equation}
The possible contribution to the gauge-boson masses from a triplet vev 
arises from the quartic part as
\begin{align}
-(\imath g_2)^2 \text{Tr}\left(\big[T^\dagger,W^\dagger\big] \big[W,T\big]\right)
&=\left(\frac{g_2}{4}\right)^2\text{Tr}\left(\left[\tau^a,\tau^b\right]\left[\tau^c,\tau^d\right]\right) T^\dagger_a W_b W_c T_d\notag\\
&=-\frac{g_2^2}{2}\left(W^a W_a T^b T^{\dagger}_b-W^a T_a^{\dagger} W^b T_b\right)\,,
\end{align}
where the $\tau^i$ are the Pauli matrices. 
The two terms of the last expression cancel each other for the neutral 
component when the triplet vev is inserted.
Hence, only the mass of the charged W boson receives an additional
 contribution. Then, the lowest-order masses of the Z and W bosons 
in the MRSSM are given as
\begin{equation}
 m_Z^2=\frac{g_1^2+g_2^2}{4}v^2,\;\quad m_W^2=\frac{g_2^2}{4}v^2+g_2^2 v_T^2\;,
\label{eq:masses}
\end{equation}
with $v=\sqrt{v_u^2+v_d^2}$.
The appearance of $v_T$ in the expression for the W boson mass is also
relevant for the definition of the weak mixing angle as the introduction of
$v_T$ spoils
the accidental custodial symmetry of the SM.

Numerically the contribution due to the triplet vev is strongly constrained
by the measurement of electroweak precision observables, 
especially $\Delta\hat\rho$
which leads to a limit of $|v_T|\lesssim3$~GeV.
A more detailed discussion of its influence on the W boson mass 
is given in sec.~\ref{sec:vt_num}.

The weak mixing angle which diagonalises the neutral vector boson mass matrix leading to the photon and Z boson is related to the gauge couplings as usual
\begin{equation}
\cmixx\equiv \cos^2(\tilde\theta_W)=\frac{g_2^2}{g_1^2+g_2^2},\;\quad \smixx=1-\cmixx\;,
\label{eq:cmixx}
\end{equation}
where we have introduced the notation $\tilde\theta_W$, $\cmix$ and $\smix$
in order to distinguish those quantitities from the ones defined in 
eq.~\eqref{eq:cmass} below.
Together with the electric charge $e$ derived from the fine-structure constant 
$\alpha=e^2/4\pi$ the weak mixing angle $\tilde\theta_W$ can be used to 
replace the gauge couplings,
\begin{equation}
g_1=\frac{e}{\cmix},\;\quad  g_2=\frac{e}{\smix}.
\label{eq:gaugecoupl}
\end{equation}
Additionally, we define the ratio of the masses of the electroweak gauge bosons as
\begin{equation}
\cmass\equiv\frac{m_W^2}{m_Z^2}=\cmixx+\frac{e^2 v_T^2}{(1-\cmixx)
m_Z^2},\;\quad \smass= 1 - \cmass =\smixx-\frac{e^2 v_T^2}{\smixx m_Z^2}\;.
\label{eq:cmass}
\end{equation}
In the limit of vanishing $v_T$ the two quantities $\smass$ and $\smixx$
coincide with each other at tree-level as in the MSSM.
For the extraction of the W boson mass from the muon decay constant it is helpful
to solve eq.~\eqref{eq:cmass} for $\smixx$
\begin{equation}
\smixx= \frac{1}{2}\left(\smass +\sqrt{\smas^4+\frac{4 e^2 v_T^2}{m_Z^2}}\right)\,
\label{eq:smixx_mw_rel}
\end{equation}
taking the physical solution 
such that 
%the previously discussed limit between 
$\smixx \to \smass$ for $v_T \to 0$ holds as required.

\subsection{Determination of the W boson mass}
\label{sec:MWdet}

The Fermi constant $G_F$ is an experimentally very precisely measured
observable that is obtained from muon decay. The comparison with the
theoretical prediction for muon decay yields a relation between the 
Fermi constant, the W boson mass, the Z boson mass and the
fine-structure constant. Therefore a common approach is to
use $G_F$ as an input in order to derive a 
prediction for the W boson mass in the SM and in BSM models.

Muons decay to almost  100\% via $\mu^-\rightarrow e^- \nu_\mu\bar\nu_e$. The Fermi model describes this interaction via 
a four-point interaction with the coupling $G_F$.
It is connected to the experimentally precisely measured 
muon lifetime $\tau$ extracted
by the MuLan experiment~\cite{Tishchenko:2012ie} via
\begin{equation}
\frac{1}{\tau_\mu}=\frac{m_\mu^5G_F^2}{192\pi^3}F\left(\frac{m_e^2}{m_\mu^2}\right)
(1+\Delta q)\;.
\label{eq:muon_width}
\end{equation}
Here, $F(m_e^2/m_\mu^2)$ collects effects of the electron mass
on the final-state phase space, and $\Delta q$ denotes the QED corrections
to the Fermi model up to two loops~\cite{vanRitbergen:1999fi,Steinhauser:1999bx,Pak:2008qt}.%
\footnote{In the past, conventionally a factor $(1+(3m_\mu^2)/(5M_W^2))$ has
often been
 inserted in eq.~\eqref{eq:muon_width} in order to take into account 
tree-level W propagator effects even though this correction term
is not part of the Fermi model.
With the enhanced accuracy of the MuLan experiment this previously numerically
negligible factor has to be taken into account on the side of the full SM calculation when using the experimentally extracted value of $G_F$ from ref.~\cite{PDG:2018}. }
Detailed expressions can be found for instance in
Chapter 10 of the Particle Data Group report~\cite{PDG:2018}.

Equating the expression in the Fermi model with the SM prediction in the
on-shell scheme yields the following relation between the Fermi constant
$G_F$, the fine-structure constant $\alpha = e^2/(4 \pi)$ in the Thomson
limit, and the pole masses (defined according to a Breit--Wigner shape with
a running width, see below) of the Z and W bosons, $M_Z$ and $M_W$, respectively,
\begin{equation}
\text{SM: } \qquad 
\frac{G_F}{\sqrt2}= \frac{e^2}{8 M^2_W \smass}\left(1+\Delta r\right)\;,
\label{eq:realdr-intro}
\end{equation}
where all higher-order contributions besides the ones appearing in 
eq.~\eqref{eq:muon_width} are contained in the quantity $\Delta r$. The same
functional relation holds also in the MSSM and the NMSSM.%
\footnote{As mentioned above, tree-level effects from the longitudinal part
of the W propagator and, for the case of an extended Higgs sector, of the
charged Higgs boson(s) are understood to be incorporated into $\Delta r$. As
effects of this kind are insignificant for our numerical analysis, we will
neglect them in the following.}
The weak mixing angle in eq.~\eqref{eq:realdr-intro} is given by 
$\smass = 1 - \MW^2/\MZ^2$ in accordance with
eq.~\eqref{eq:cmass}.

In the MRSSM the relation of eq.~\eqref{eq:realdr-intro} gets modified
because of the contribution of the triplet vev $v_T$
entering at lowest order,
\begin{equation}
\text{MRSSM: } \qquad 
\frac{G_F}{\sqrt2}= \frac{e^2}{8 M^2_W \smixx}\left(1+\Dhr\right)
\label{eq:dr-intro}
\end{equation}
where $\smixx$ originates from the gauge coupling $g_2$, see 
eq.~\eqref{eq:gaugecoupl} and eq.~\eqref{eq:cmass}. We use the notation 
$\Dhr$ for the higher-order contributions in the MRSSM,
where eq.~\eqref{eq:dr-intro} is expressed in terms of the quantities 
$\MW$, $\MZ$, $\alpha$ defined in the on-shell scheme
as in eq.~\eqref{eq:realdr-intro}.
Inserting eq.~\eqref{eq:smixx_mw_rel} into eq.~\eqref{eq:dr-intro} and solving
for $M_W^2$ yields
\begin{equation}
M_W^2= M_Z^2 \left(\frac{1}{2}+\sqrt{\frac{1}{4}-\frac{\alpha\pi}{\sqrt2 G_F 
M_Z^2}(1+\Dhr-4\sqrt2 G_F v_T^2)}\right)\cdot\left(\frac{1}{1 - \frac{4\sqrt2 G_F v_T^2}{1+\Dhr }}\right)\,.
\label{eq:mw_dr}
\end{equation}
As $\Dhr$ itself depends on $M_W$ it is technically most convenient to
determine $\MW$ numerically through an iteration of this relation.
In the limit $v_T \to 0$ the result of eq.~\eqref{eq:mw_dr} yields the
well-known expression
\begin{equation}
M_W^2\Bigr|_{v_T \to 0} 
= M_Z^2 \left(\frac{1}{2}+\sqrt{\frac{1}{4}-\frac{\alpha\pi}{\sqrt2 G_F 
M_Z^2}(1+\Dr)}\right)\,
\label{eq:mw_realdr}
\end{equation}
that is valid in the SM, since $\Dhr$ coincides with the usual definition of
$\Dr$ in this case.

As mentioned above, in the previous calculation~\cite{PD1} for the MRSSM 
a mixed scheme was used where only the gauge-boson masses $\MW$ and $\MZ$
are on-shell quantities.
For completeness we provide a short description of this scheme.
Following refs.~\cite{Degrassi:1990tu,Degrassi:2014sxa}, 
the running electromagnetic coupling $\hat \alpha_{\MS /\DR}$
and the running mixing angle $\hat s^2_{W,\MS/\DR}$ are used in this scheme
together with the on-shell definitions of the masses.
Here, $\MS/\DR$ denotes renormalisation via modified minimal subtraction 
either in dimensional regularisation or reduction.
Higher-order corrections are collected in several parameters that
incorporate a resummation of certain reducible higher-order contributions,
in particular
\begin{equation}
\hat\alpha_{\MS /\DR} = \frac{\alpha}{1-\Delta \hat \alpha}\;, \quad
\hat c^2_{W,\MS /\DR}= \frac{M_W^2}{M_Z^2\hat\rho}\;, \quad
\hat s^2_{W,\MS/\DR} = 1 - \hat c^2_{W,\MS /\DR}\;, 
\label{eq:hatparams}
\end{equation}
where 
\begin{align}
\Delta\hat\alpha &= \hat\Pi^{AA}(0)\;, \quad 
\hat\Pi^{AA}(0)\equiv
\left.\frac{\partial \hat\Sigma^{AA}_T(p^2)}{\partial p^2}\right|_{p^2=0}\;,
\quad
\hat\rho = \frac{1}{1 - \Delta\hat\rho_0 - \Delta\hat\rho} \; ,
\quad\notag\\
\Delta\hat\rho_0 &= \frac{4 v_T^2}{v_T^2 + v^2}\; , \quad
\Delta\hat\rho = \Re \left(
\frac{\hat\Sigma_T^{ZZ}(M^2_{Z})}{M_Z^2} -
\frac{\hat\Sigma_T^{WW}(M^2_{W})}{M_W^2}\right) \; .
\label{eq:hatrho}
\end{align}
Here $\Delta\hat\rho_0$ contains the tree-level shift arising from the
triplet vev, while $\Delta\hat\rho_0$ contains the higher-order corrections
including all MRSSM one-loop and leading SM two-loop effects, 
$\hat\Sigma_T$ denotes the transverse part of the 
$\DR$-renormalised self-energies of the gauge
bosons, and $\Re$ indicates the real part.
Analogously to eq.~\eqref{eq:dr-intro} one can write
\begin{equation}
\text{MRSSM: } \qquad 
\frac{G_F}{\sqrt2}= \frac{4\pi \hat \alpha_{\MS /\DR}}{8 M^2_W 
\hat s^2_{W,\MS/\DR}}\left(1+\Delta\hat r_W\right)\; ,
\label{eq:hatDeltar}
\end{equation}
where the quantities defined in eqs.~\eqref{eq:hatparams}, 
\eqref{eq:hatrho} have been used, and 
$\Delta\hat r_W$ contains additonal 
higher-order corrections.
The described approach has been used in previous work for the MRSSM and is the one
implemented in several MSSM and NMSSM mass spectrum generators following
refs.~\cite{Chankowski:1994ua,BPMZ}.
For an approach in a pure minimal subtraction scheme see
ref.~\cite{Martin:2015lxa}. 

In the following, we discuss the details of the on-shell calculation in the
MRSSM according to eq.~\eqref{eq:dr-intro}, making use of the
state-of-the-art prediction of the SM part,
and compare it to previously obtained results. 

\section{\Dhr\ in the MRSSM}
\label{sec:dr}
For our on-shell calculation of the W boson mass in the MRSSM we include all 
MRSSM SUSY effects at one-loop level and all known higher-order contributions
of SM type. This follows the approach taken for the 
MSSM~\cite{Stal:2015zca,Heinemeyer:2006px} and the NMSSM~\cite{Stal:2015zca}.
\subsection{One-loop corrections}
\subsubsection{General contributions}
At the one-loop level, $\Dhr^{(\alpha)}$ receives contributions from the W
boson self-energy, from vertex and box corrections, as well as from the
corresponding counterterm diagrams. 
It can be written as
\begin{equation}
\Dhr^{(\alpha)}=\frac{\Sigma_T^{WW}(0) -\delta M_W^2}{\MW^2}
+\text{(Vertex and Box)}
+\frac{1}{2}\left(\delta Z^e_L+\delta Z^\mu_L+\delta Z^{\nu_e}_L+\delta Z^{\nu_\mu}_L\right) +2\delta e - \frac{\delta \smixx}{\smixx}\;.
\label{eq:dr-start}
\end{equation}
The field renormalisation of the W boson drops out as the field only appears
internally. %
The expressions for the MRSSM vertex and box contributions are given
in the appendix of ref.~\cite{PD1}.

Using the on-shell scheme, we fix the renormalisation constants of the 
gauge-boson masses as
\begin{equation}
M^2_{W/Z,0}=M^2_{W/Z}+\delta M^2_{W/Z}\;,\quad\delta M^2_{W/Z}=
\Re \Sigma_T^{WW/ZZ}(M^2_{W/Z})\;,
\end{equation}
where $\Sigma_T^{WW/ZZ}(p^2)$ is the transverse part of the unrenormalised
gauge-boson self-energy taken at momentum $p^2$, and as before $\Re$ denotes
the real part.
The field renormalisation constant of a massless left-handed lepton is 
\begin{equation}
l_{L,0}=\left(1+\frac{1}{2}\delta Z^l_L\right)l_L\;,\quad
\delta Z^l_L=-\Sigma_L^l(0)\;.
\end{equation}
The electric charge is renormalised as 
\begin{equation}
\begin{split}
e_0&=e(1+\delta e)\;, \quad \delta e = \frac{1}{2} \Pi^{AA}(0)+ 
\frac{\smix}{\cmix}
\frac{\Sigma^{AZ}_T(0)}{M_Z^2}\;,\;\Pi^{AA}(0)\equiv
\left.\frac{\partial \Sigma^{AA}_T(p^2)}{\partial p^2}\right|_{p^2=0}\,,\\
&\Delta\alpha^{(5)}_{\text{had}}(M_Z^2) = \left(\left.\Pi^{AA}(0)-\Re\Pi^{AA}(M_Z^2)\right)\right|_{\text{light quarks}}\,,
\end{split}
\end{equation}
where the quantity $\Delta\alpha^{(5)}_{\text{had}}(M_Z^2)$ is extracted
from experimental data and accounts for the contributions 
of the five light quark flavours. 

For the renormalisation of the weak mixing angle \smixx\ the appearance of $v_T$
leads to differences compared to the SM and the MSSM.
The parameter $\smass$ is renormalised with the renormalisation constants of
the gauge-boson masses
\begin{equation}
{\smass}_{,0}=\smass+\delta\smass\;,\quad\frac{\delta \smass}{\smass}= 
\frac{\cmass}{\smass}\Re\left(\frac{\Sigma_T^{ZZ}(M^2_{Z})}{M^2_{Z}} - 
\frac{\Sigma_T^{WW}(M^2_{W})}{M^2_{W}}\right)\;.
\end{equation}
The renormalisation constant of the weak mixing angle 
$\smixx$ can be expressed in terms of $\delta\smass$, $\delta e$, $\delta v_T$ 
and $\delta\MZ^2$ using the relation between $\smixx$ and 
$\smass$ given in eq.~\eqref{eq:cmass},
\begin{equation}
\smixx{}_{,0}=\smixx+\delta\smixx\;,\quad\frac{\delta \smixx}{\smixx}=
\left\{\frac{\delta \smass}{\smass}
-
\frac{\smass-\smixx}{\smass} \left[2\left(\delta e + \frac{\delta v_T}{v_T}
\right) - \frac{\delta M_Z^2}{M_Z^2}\right]\right\}\left(\frac{\smass}{2\smixx-\smass}\right)\;.
\label{eq:smixx_renorm1}
\end{equation}
Expressing $\delta \smixx$ by the self-energies of the vector bosons
gives
\begin{equation}
\begin{split}
\frac{\delta \smixx}{\smixx}=&\frac{\cmass}{2\smixx-\smass}\Re\left(
\frac{\Sigma_T^{ZZ}(M^2_{Z})}{M^2_{Z}}-\frac{\Sigma_T^{WW}(M^2_{W})}{M^2_{W}}\right) \\
&+ \frac{4 \pi\alpha v_T^2}{\smixx(2\smixx-\smass) M_Z^2}\left( \Pi^{AA}(0) + 2 \frac{\smix}{\cmix} 
\frac{\Sigma^{AZ}_T(0)}{M_Z^2} -\Re \frac{\Sigma_T^{ZZ}(M^2_{Z})}{M_Z^2}+
2\frac{\delta v_T}{v_T}\right)\, .
\label{eq:smixx_renorm}
\end{split}
\end{equation}
If the triplet vev was absent, the on-shell renormalisation of the electroweak 
parameters would be the same as in the SM and the MSSM, and $\delta\smixx$ and
$\delta\smass$ would coincide. It is important to note in this context that
in our renormalisation prescription $\smixx$ is treated as a dependent
parameter as specified in eq.~\eqref{eq:smixx_renorm1}.
The renormalisation of the triplet vev is described in the following section.
This prescription for the weak mixing angle ensures that the contributions to 
$\delta \smixx/\smixx$ incorporate the typical quadratic dependence
on $m_t$ that is induced by the contribution of the top/bottom doublet to
the $\rho$ parameter at one-loop order. This behaviour was found to be
absent in renormalisation schemes where the weak mixing angle is treated as
an independent parameter that is fixed as a process-specific effective
parameter $\sin^2\theta_{\text{eff}}$, 
see refs.~\cite{Blank:1997qa,Czakon:1999ha,Chen:2005jx}. 

\subsubsection{Renormalisation of $v_T$}
\label{sec:vt}
The triplet vev $v_T$ is an additional parameter of the electroweak sector in the 
MRSSM compared to the MSSM. As it appears in the lowest-order 
relation between the muon 
decay constant and $M_W$, eq.~\eqref{eq:mw_dr},%
\footnote{It also appears in the lowest-order relation
between $\MW$ and the weak mixing angle, eq.~\eqref{eq:smixx_mw_rel}.}
its renormalisation is required for the prediction of $M_W$.
In principle, in the MRSSM one could instead have used $\MW$ as an experimental
input parameter in order to determine $v_T$ via 
eqs.~\eqref{eq:masses}--\eqref{eq:smixx_mw_rel}. Among other drawbacks, in
such an approach the SM limit of the MRSSM, which involves $v_T \to 0$,
could not be carried out. Instead, we prefer to keep $\MW$ as an observable
that can be predicted, using in particular $G_F$ as an experimental input,
and compared to the experimental result as it is the
case in the SM and the MSSM.

For the calculation of $M_W$ performed in this work we renormalise $v_T$ 
such that 
\begin{equation}
\delta v_T\overset{!}{=}\kappa \left(\frac{2}{4-D}-\gamma_E+\log 4\pi\right)\,.
\label{eq:vt-renorm}
\end{equation}
Accordingly, $\delta v_T$ only contains the  divergent contribution
(in the modified minimal subtraction scheme)
with a prefactor
$\kappa$, where the correct choice of the latter ensures that the
renormalised quantities are finite. 
The value of $v_T$ is taken as input at the SUSY scale $m_{\text{SUSY}}$.
The triplet vev is the only BSM parameter entering the tree-level expression
of eq.~\eqref{eq:masses}. All other BSM parameters of the MRSSM
only enter the loop contributions and do not need to be renormalised at the one-loop level.
As $v_T$ is a parameter of the electroweak potential a comment on the tadpole
conditions is in order.
The tree-level tadpoles are given as
\begin{align} 
{t_{d}} =&v_d [\textstyle{\frac{1}{8}}\left(g_1^2+g_2^2\right) \left(v_d^2-v_u^2\right)- g_1 M_B^D v_S + g_2 M_W^D v_T+m_{H_d}^2+ (\mud{+})^2]- v_u B_\mu\;, \notag\\
{t_{u}}=&v_u [\textstyle{\frac{1}{8}}\left(g_1^2+g_2^2\right) \left(v_u^2-v_d^2\right)
+g_1 M_B^D v_S- g_2 M_W^D v_T+m_{H_u}^2+ (\muu{-})^2]- v_d B_\mu\;, \notag\\
{t_{T}}=& \textstyle{\frac{1}{2}} g_2
M_W^D \left(v_d^2-v_u^2\right)+
\textstyle{\frac{1}{2}}\left(\Lambda_d v_d^2  \mud{+} -\Lambda_u v_u^2 \muu{-} \right)
+4 (M_W^D)^2 v_T+ m_T^2 v_T\;, \notag\\
{t_{S}}=& \textstyle{\frac{1}{2}}g_1
M_B^D \left(v_u^2-v_d^2\right)+
\textstyle{\frac{1}{\sqrt{2}}}\left( \lambda_d v_d^2 \mud{+}
+ \lambda_u v_u^2 \muu{-}\right) +4 (M_B^D)^2 v_S+ m_S^2 v_S\;,
\label{eq:tadp}
\end{align}
and one may trade one model parameter for each of the tadpoles.
The numerical values of those parameters are fixed implicitly by the conditions $t_i\equiv0$
so that they now are expressed as functions of all the other input parameters of the model.
In our calculation of $M_W$ we choose the set of $m_{H_u}^2$, $m_{H_d}^2$,
$m_{T}^2$ and $v_S$ as dependent parameters.
Choosing instead $v_T$ as a parameter to be fixed by the tadpole equations would require a renormalisation of all parameters in these relations. 
This would lead to a finite counterterm 
$\delta v_T$ which would have a complicated form but would be expected to
have a numerically 
very small impact 
on the prediction of $M_W$
as it is suppressed in eq.~\eqref{eq:smixx_renorm} by a prefactor of $v_T^2/M_Z^2$.

Our calculation for the $M_W$ prediction in the MRSSM is embedded into the framework of
\texttt{SARAH}/\texttt{SPheno} described in section~\ref{sec:implementation}
below.
There, we use a different set of parameters derived from solving the tadpole
equations,
namely $v_T$ is treated as an output there and $m_T^2$ as input. The value for $v_T$
calculated by the \texttt{SARAH}/\texttt{SPheno} routines is then passed to our
implementation of the $M_W$ calculation as an input.
This is necessary as $v_T$ is much smaller than $m_{\text{SUSY}}$ for realistic parameter points.
Therefore, small variations in $v_T$ during the required iteration might have strong effects
on the BSM mass spectrum and may lead to numerical instabilities.
For example, at tree-level a chosen $v_T$ value might lead to physical Higgs states
while tachyonic states might appear after adding the loop corrections to the
mass matrices and tadpole equations. This is circumvented by using in a
first step $m_T^2$ as input and keeping it positive.
In the following we describe the treatment of the tadpole conditions 
in the \texttt{SARAH}/\texttt{SPheno} framework 
in more detail and illustrate that the definition that we have chosen
for $v_T$ ensures that it
is suitable as input for
our calculation of $M_W$ satisfying eq.~\eqref{eq:vt-renorm}.  

In the \texttt{SARAH}/\texttt{SPheno} framework free parameters are renormalised in the $\DR$ scheme, and the masses of the SUSY states are calculated to a considered loop order.
The tadpoles are an exception in this context as they are renormalised
relying on on-shell conditions. Hence, their counterterms cancel the tadpole
diagrams, and the tadpole contributions do not need to be considered as subdiagrams
of other loop diagrams. 
This means that the bare tadpole $t_{0,i}$ is given as
\begin{equation}
t_{0,i}=\hat t_i+\delta t_i\;, \quad\delta t_i = - \Gamma^i(0)\;,
\end{equation}
where $\Gamma^i(0)$ are the unrenormalised one-loop one-point functions.
The renormalised tadpoles $\hat t_i$ are required to vanish, 
corresponding to the minimum of the effective potential.
The parameters chosen as dependent parameters via the tadpole conditions for the \texttt{SARAH}/\texttt{SPheno} mass spectrum calculation are
$\tilde m_{H_u}^2$, $\tilde m_{H_d}^2$, $\tilde v_{T}$ and $\tilde v_S$. As the tadpoles are renormalised
via on-shell conditions, the renormalised dependent parameters respect the
tree-level relations
while the counterterms of these parameters ($\delta\tilde m_{H_u}^2$,
$\delta\tilde m_{H_d}^2$, $\delta \tilde v_{T}$, $\delta\tilde v_S$) have to
contain finite parts. Therefore, considering
only the finite part of all appearing quantities (i.e., counterterms that
only contain a divergent contribution have been dropped)
the finite parts of the counterterms are given implicitly via the following
relations:
\begin{align}
v_d[(\textstyle{\frac{1}{4}}\Lambda_d^2 \tilde v_T+\textstyle{\frac{1}{2}}\Lambda_d \mu_d+g_2 M_W^D)
\delta \tilde v_{T}+(\textstyle{\frac{1}{2}}\lambda_d^2\delta\tilde v_S+\lambda_d \mu_d- g_1 M_B^D)\delta\tilde v_S+
\delta\tilde m_{H_d}^2]&=-\Gamma^d(0)\;, \notag\\
v_u[(\textstyle{\frac{1}{4}}\Lambda_d^2 \tilde v_T-\textstyle{\frac{1}{2}}\Lambda_u \mu_u-g_2 M_W^D)
\delta \tilde v_{T}+(\textstyle{\frac{1}{2}}\lambda_u^2\delta\tilde v_S+\lambda_u \mu_u+g_1 M_B^D)\delta\tilde v_S
+\delta\tilde m_{H_u}^2]&=-\Gamma^u(0)\;, \notag\\
\textstyle{\frac{1}{2\sqrt{2}}}(\lambda_d\Lambda_d v_d^2-\lambda_u\Lambda_u v_u^2)\delta \tilde v_{S}+[m_T^2+4 (M_W^D)^2+\textstyle{\frac{1}{4}}(\Lambda_d^2v_d^2+\Lambda_u^2 v_u^2)]\delta \tilde v_{T}&=-\Gamma^T(0)\;, \notag\\
\textstyle{\frac{1}{2\sqrt{2}}}(\lambda_d\Lambda_d v_d^2-\lambda_u\Lambda_u v_u^2)\delta \tilde v_{T}+[m_S^2+4 (M_B^D)^2+\textstyle{\frac{1}{2}}(\lambda_d^2v_d^2+\lambda_u^2 v_u^2)]\delta \tilde v_{S}&=-\Gamma^S(0)\;.
\label{eq:tadploop}
\end{align}
Note that the terms containing $\lambda_{d,u}$ and $\Lambda_{d,u}$ arise
from the terms involving $\mud{\pm}$ and $\muu{\pm}$ in eq.~\eqref{eq:tadp}.

In the loop calculation one can then define parameters 
$m_{H_u}^{2,\text{loop}}$, $m_{H_d}^{2,\text{loop}}$, $ v_{T}^\text{loop}$
and $ v_S^\text{loop}$ as sum of the tree-level parameter and the finite
part of the counterterm. 
For the triplet vev we choose (as above, a purely divergent counterterm is
dropped)
\begin{equation}
 v_{T}^\text{loop}= \tilde v_T + \delta \tilde v_{T}|_{\text{finite}}\;.
\end{equation}
%Following this convention, the RGEs of
%the vevs can be derived then from the general expressions given in refs.~\cite{Sperling:2013eva,Sperling:%2013xqa}. 
This definition of $v_{T}^\text{loop}$ 
corresponds to a renormalisation of $v_T$ with a purely
divergent counterterm as 
required in eq.~\eqref{eq:vt-renorm}.
Therefore, $v_T^\text{loop}$ as an output of this procedure is a suitable input for the our calculation of the $M_W$ prediction.
This change in parameterisation leads to a numerical difference between
the value of $m_T^2$ used for the $M_W$ calculation on the one hand and for
the derivation of the loop-corrected SUSY mass spectrum on the other.
As we consider the SUSY corrections only up to the one-loop level
for $\Dhr$ the one-loop shift of
the model parameter $m_T^2$ formally leads to a higher-order effect 
that is beyond the considered order for the SUSY loop corrections.
Solving eq.~\eqref{eq:tadploop} for $\delta \tilde v_{T}|_{\text{finite}}$
yields
\begin{align}
\delta &\tilde v_{T}|_{\text{finite}}=\\
&\frac{[m_S^2+4 (M_B^D)^2+\textstyle{\frac{1}{2}}(\lambda_d^2v_d^2+\lambda_u^2 v_u^2)]\Gamma^T(0)|_\text{finite}-\textstyle{\frac{1}{2\sqrt{2}}}(\lambda_d\Lambda_d v_d^2-\lambda_u\Lambda_u v_u^2)\Gamma^S(0)|_\text{finite}}{\textstyle{\frac{1}{8}}(\lambda_d\Lambda_d v_d^2-\lambda_u\Lambda_u v_u^2)^2-[m_T^2+4 (M_W^D)^2+\textstyle{\frac{1}{4}}(\Lambda_d^2v_d^2+\Lambda_u^2 v_u^2)][m_S^2+4 (M_B^D)^2+\textstyle{\frac{1}{2}}(\lambda_d^2v_d^2+\lambda_u^2 v_u^2)]}.
\end{align}
The divergent part of $\delta \tilde v_{T}$ on the other hand
can be compared to the general expressions given in ref.~\cite{Sperling:2013eva} and we find agreement.

A compact expression for $v_T$ can be derived by solving the third equation of
\eqref{eq:tadp} for it. At tree-level, we get the following 
\begin{equation}
v_T=\frac{(\Lambda_u\muu{0}+g_2 M_W^D ) v_u^2 - (\Lambda_d\mud{0}+g_2 M_W^D ) v_d^2 
}{2\left(m_T^2+4
(M_W^D)^2\right)+\frac{1}{2}\left(\Lambda_d^2v_d^2+\Lambda_u^2v_u^2\right)}\;.
\label{eq:vt}
\end{equation}%+\frac{\Lambda_d^2\v_d^2+\Lambda_u^2\v_u^2}{2}
The magnitude of $v_T$ can be affected in several ways.
On the one hand it can become small as a consequence of
large SUSY mass scales appearing in
the denominator. Here, the combination $m_T^2+4 (M_W^D)^2$ is the squared tree-level
mass matrix entry for the CP-even Higgs triplet.
On the other hand, the numerator can become small. 
For $\tan\beta>1$ the term proportional to $v_u^2$ dominates.
If then $g_2 M_W^D$ is numerically close to $-\Lambda_u\muu{0}$ a partial
cancellation is possible leading to a reduced value for $v_T$.
\footnote{It should be noted that the expressions in parenthesis in the numerator
also appear in the Higgs mass matrix elements mixing the doublets and triplet. 
Therefore, if the triplet scalar mass parameter is not large,
a certain tuning is necessary to reduce the admixture
of the triplet Higgs component with the state at
125~GeV in order to ensure that the latter is sufficiently SM-like.
Such a tuning would at the same time reduce the numerical value 
of $v_T$. In the numerical scenarios studied in this paper, however, such a
tuning does not occur since in our numerical analysis below
the triplet mass is always much larger than the mass of the SM-like state in the
parameter regions where
the $M_W$ prediction is close to the experimental measurement.}

In ref.~\cite{Chankowski:2006hs} the influence of the triplet vev on the decoupling behaviour 
in a model where the Standard Model is extended by a real triplet was studied. 
It was found
that non-decoupling behaviour exists when the triplet mass parameter
approaches large
values while the triplet-doublet-doublet trilinear coupling also grows.
While a detailed investigation of this issue in the MRSSM would go beyond the 
scope of the present paper, we note that 
studying the numerical one-loop contributions to the triplet tadpole
$\Gamma^T$ 
we do not find a non-decoupling effect for a comparable limit. Hence,
this effect does not appear in our numerical analysis, see also the
discussion in section~\ref{sec:gen_results} where we compare the MRSSM
prediction with the one of the SM for the case where the SUSY mass scale is
made large. Our results regarding the decoupling behaviour of the MRSSM
contributions can be qualitatively understood in the following way. 
In an effective field theory analysis of the decoupling behaviour 
one would study the matching of the MRSSM
to an effective SM+triplet model. There, the tree-level matching conditions
would fix the quartic triplet coupling to zero as it does not appear due to
R-symmetry in the MRSSM.
This affects the number of free parameters of the effective model, 
preventing the occurrence of non-decoupling behaviour.
However, if one-loop matching conditions for the quartic coupling
were taken into account these contributions could in principle give rise 
to a non-decoupling effect. 
As those contributions would correspond to a two-loop effect in the MRSSM 
for the calculation of $M_W$ they are 
outside of the scope of the present work, and we leave the investigation of
contributions of this kind to further study.

\subsection{Higher-order contributions}

For a reliable prediction of $\MW$ in a BSM model 
it is crucial to take into account
SM-type corrections beyond one-loop order. Only upon the incorporation of
the relevant two-loop and even higher-order SM-type loop contributions it is
possible to recover the state-of-the-art SM prediction within the current
experimental and theoretical uncertainties in the appropriate
limit of the BSM parameters, see e.g.\ the discussion in
ref.~\cite{Stal:2015zca}. In our predictions we incorporate the complete
two-loop and the numerically relevant higher-order SM-type
contributions. 

A further important issue in this context is the precise definition of the
gauge-boson masses according to a Breit--Wigner resonance shape with running
or fixed width. While the difference between the two prescriptions
formally corresponds to an electroweak two-loop effect, numerically the
associated shifts are about 27~MeV for $\MW$ and 34~MeV for $\MZ$.

\subsubsection{Breit--Wigner shape}

The definition of the masses of unstable particles according to the real part of 
the complex pole of their propagator is gauge-independent also beyond
one-loop order, while the definition according to the real pole is not. 
Expanding the propagator around the complex pole leads to a Breit--Wigner shape
with fixed width (f.w.). Experimentally, the gauge boson masses are
extracted, by definition, from a
Breit--Wigner shape with a running width (r.w.). Hence, it is necessary to translate
from the internally used fixed-width mass $M_W^{\text{f.w.}}$ to the
running-width mass $M_W^{\text{r.w.}}$ at the end of the calculation
\begin{equation}
M_W^{\text{r.w.}}=M_W^{\text{f.w.}}+\frac{\Gamma^2_W}{2 M_W^{\text{r.w.}}}\;,
\end{equation}
where for the decay width of the W boson, $\Gamma_W$, we use the theoretical 
prediction parametrised by $G_F$ including first order QCD corrections,
\begin{equation}
\Gamma_W= \frac{3 (M_W^{\text{r.w.}})^3 G_F}{2\sqrt{2}\pi}
\left(1 + \frac{2\alpha_s}{3\pi}\right)\;.
\end{equation}
For $\MZ$, which is an input parameter in the prediction
for $\MW$, the conversion from the running-width to the fixed-width definition
is carried out in the first step of the calculation. 

In the following, the labels f.w.\ and r.w.\ will not be displayed
as the fixed-width gauge boson masses only appear internally in the calculation. 
If not stated differently, the parameters $M_W$ and $M_Z$ always refer 
to the definition of the gauge boson masses according to a
Breit--Wigner shape with running width.

\subsubsection{Higher-order SM-type contributions}
\label{sec:higher-orderSM}

The part of $\Delta \hat r$ beyond the one-loop level contains all known
SM-type contributions.
Explicitly, as in ref.~\cite{Stal:2015zca} we write
\begin{equation}
\Dhr^{\text{MRSSM}}=\Dhr^{\text{MRSSM}(\alpha)}+\Dhr^{\text{MRSSM(h.o.)}},
\end{equation}
where $\Dhr^{\text{MRSSM}(\alpha)}$ contains all one-loop corrections from the 
different sectors:
\begin{equation}
\Dhr^{\text{MRSSM}(\alpha)}=\Dhr^{(\alpha)}_{\text{fermion}}+\Dhr^{(\alpha)}_{\text{gauge-boson/Higgs}}+\Dhr^{(\alpha)}_{\text{sfermion}}+\Dhr^{(\alpha)}_{\text{chargino/neutralino}}
+\Dhr^{(\alpha)}_{\text{R-Higgs}}\;.
\end{equation}
The term $\Dhr^{\text{MRSSM(h.o.)}}$ denotes all higher-order corrections,
where for this work we restrict ourselves to the the state-of-the-art
SM-type corrections:%
\footnote{The MSSM-like $\mathcal{O}(\alpha\alpha_s)$ two-loop 
corrections~\cite{Djouadi:1996pa,Djouadi:1998sq,Weiglein:1998ju,Heinemeyer:2004gx}
cannot be taken over to the MRSSM case since 
the Dirac nature of the gluino modifies the gluino and mass-shift
contributions,
in contrast to the case of the NMSSM~\cite{Stal:2015zca}.
The higgsino contributions of $\mathcal{O}(\alpha_t^2,\alpha_t\alpha_b,\alpha_b^2)$ are similarly affected.
}
%\htC{The reducible contributions from eq.~(38) of 
%ref.~\cite{Stal:2015zca} have not been included.}
\begin{equation}
\Dhr^{\text{MRSSM(h.o.)}}=\Dr^{\text{SM(h.o.)}}.
\end{equation}
This includes QCD two-loop,
$\Dr^{(\alpha\alpha_s)}$~\cite{Djouadi:1987gn,Djouadi:1987di,Kniehl:1989yc,
Halzen:1990je,Kniehl:1991gu,Kniehl:1992dx},
and three-loop corrections, 
$\Dr^{(\alpha\alpha_s^2)}$~\cite{Avdeev:1994db,Chetyrkin:1995ix,Chetyrkin:1995js,
Chetyrkin:1996cf},  electroweak two-loop fermionic and bosonic corrections,
 $\Dr^{(\alpha^2)}_{\text{ferm}}$ and $\Dr^{(\alpha^2)}_{\text{bos}}$~\cite{Freitas:2000gg,Freitas:2002ja,Awramik:2002wn,
Awramik:2003ee,Onishchenko:2002ve,Awramik:2002vu}, as well as
leading mixed QCD-electroweak three-loop, purely electroweak three-loop
and QCD four-loop corrections, $\Dr^{(G_F^2 m_t^4\alpha_s )}$, 
$\Dr^{(G_F^3 m_t^6)}$~\cite{Boughezal:2004ef,Boughezal:2006xk,Faisst:2003px,
vanderBij:2000cg} and $\Dr^{(G_F m_t^2\alpha_s^3 )}$~\cite{Chetyrkin:2006bj,
Schroder:2005db}.

The full electroweak two-loop corrections in the SM, 
$\Dr^{(\alpha^2)} = \Dr^{(\alpha^2)}_{\text{ferm}} + 
\Dr^{(\alpha^2)}_{\text{bos}}$~\cite{Freitas:2000gg,Freitas:2002ja,Awramik:2002wn,
Awramik:2003ee,Onishchenko:2002ve,Awramik:2002vu}, 
for which numerical integrations of two-loop integrals with non-vanishing
external momentum are required, are implemented using the simple fit formula
given in ref.~\cite{Awramik:2006uz}. In this way 
$\Dhr^{\text{MRSSM}}(\MW)$ can be evaluated at the correct value of $\MW$ at
each step of the iterative evaluation of eq.~\eqref{eq:mw_dr},
see ref.~\cite{Stal:2015zca} for more details.
For the implementation of the SM corrections the contributions given in table 1 of
ref.~\cite{Stal:2015zca} could be reproduced taking the same input parameters.

\section{Implementation and estimate of remaining theoretical
uncertainties}
\label{sec:implementation}

As for the previous work on the MRSSM~\cite{PD1,PD2,PD3,PD4,Diss}, our
prediction for $\MW$ is embedded 
%the calculations relied on a mass spectrum generator derived 
in the framework of \texttt{SARAH}/\texttt{SPheno}~\cite{Staub:2008uz,Staub:2009bi,
Staub:2010jh,Porod:2011nf,Staub:2012pb,Staub:2013tta,Goodsell:2014bna,
Goodsell:2015ira}, see also the discussion in section~\ref{sec:vt}.
%This allowed for the calculation of the complete SUSY mass 
%spectrum following ref.~\cite{BPMZ}. The prediction for $M_W$ has been derived 
%together with the determination of the weak mixing angle $\hat s^2_{W,\DR}$ in 
%a mixed on-shell/$\DR$ scheme.
%In the following, a short review of the implementation is given. 
%Then, the changes to the code incorporating the on-shell $M_W$ calculation 
%are described.
We use \texttt{SARAH}-4.12.3 and \texttt{SPheno}-4.0.3 for this work. 
The complete calculation of the SUSY pole masses at the 
one-loop level in this framework is done in the \DR\ renormalisation scheme.
For this purpose no counterterms are calculated explicitly, but rather an implicit 
renormalisation is performed, such that divergences (including terms of $\gamma_E$
and $\log 4\pi$) are dropped keeping only the finite part of loop functions. 
The input parameters of the calculation are the SUSY parameters at the SUSY
mass scale $m_{\text{SUSY}}$
and $\alpha$, $G_F$, $M_Z$ as well as the quark and lepton masses.
As explained in section~\ref{sec:vt}, within the
\texttt{SARAH}/\texttt{SPheno} framework the tadpole equations of the MRSSM 
including the relevant higher-order
corrections are solved to obtain $m_{H_u}^2$, $m_{H_d}^2$, $v_S$ and $v_T$
at $m_{\text{SUSY}}$ in terms of the other model parameters.
In order to determine the SUSY mass spectrum 
%consistently 
two-loop renormalisation
group equations are used to run between
the electroweak scale and the SUSY scale. At the SUSY scale the pole masses of
the superpartners are calculated at the one-loop order. The Higgs boson
masses are
calculated including also two-loop effects.%
\footnote{\texttt{SARAH}/\texttt{SPheno} in principle
supports EFT matching to the SM for the calculation of the mass of the
SM-like Higgs boson. However, 
%taking triplet vev effects into account for the matching 
%procedure and including the on-shell $M_W$ calculation to derive 
%$\hat s^2_{W, \DR}$ in this approach is not straightforward.
this would require specific adjustments for the case of the MRSSM.}
As discussed in section~\ref{sec:MWdet}, in the previous calculation~\cite{PD1} 
for the MRSSM the predictions for $M_W$ and also the weak mixing angle 
$\hat s^2_{W,\DR}$ were derived from 
eqs.~\eqref{eq:hatparams}--\eqref{eq:hatDeltar}
using a mixed on-shell/$\DR$ calculation following ref.~\cite{BPMZ}.

The prediction for $M_W$ presented in this work, which employs the on-shell
scheme and incorporates the state-of-the-art SM-type contributions, has been
integrated into the described framework by extending the generated 
\texttt{SARAH}/\texttt{SPheno} output by additional 
routines implementing the calculation described in the previous sections.
For this part of the calculation, all SM parameters are renormalised on-shell 
while the SUSY masses are obtained from the input parameters via
lowest-order relations.
%we take the tree-level masses calculated from the 
As described in section~\ref{sec:vt}, $v_T$ is renormalised as $\DR$ parameter
and taken at the SUSY scale. 
No renormalisation of the other SUSY parameters is required as they enter only 
at the one-loop level. 
This ensures the validity of symmetry relations between the SUSY parameters.
%This allows for consistent numerical cancellations 
%between different contributions to the gauge boson self-energies. 
The obtained fully analytical expressions for the one-loop parts are
combined with the state-of-the-art higher-order corrections of SM-type
as described in section~\ref{sec:higher-orderSM}. 
%including the
%electroweak two-loop contributions for $\Dr^{\alpha^2,SM}(M_W)$ of
%ref.~\cite{Awramik:2006uz} as fit formula allowing for a consistent
%treatment of the MRSSM $M_W$ prediction in the SM part of $\Delta r$.
%To arrive at a consistent result for the calculation an iteration is 
%required so that numerical convergence is ensured.
The evaluation of eq.~\eqref{eq:mw_dr} is carried out iteratively until
numerical convergence is obtained. The prediction for the 
W~boson mass obtained in this way is then employed to 
extract $\hat s^2_{W,\DR}$ and calculate the $\DR$ gauge couplings 
using the parameter $\hat\rho$ according to eq.~\eqref{eq:hatrho}. These
quantities are used as part of the SUSY mass spectrum calculation.

\medskip

%\subsection{Theoretical uncertainties}
Concerning the remaining theoretical uncertainties of the prediction for the
W~boson mass in the MRSSM, one needs to account for theoretical
uncertainties that are induced by the experimental errors of the input
parameters as well as for uncertainties from unknown higher-order
corrections.
%There are several sources of uncertainties entering the calculation. 
%Following the arguments in previous calculations~\cite{Heinemeyer:2006px,Stal:2015zca} 
%We summarise them here.
%On the one hand, each of the input parameters has an uncertainty from the
%measurement.
Among the experimental errors of the input parameters the one associated
with the top-quark mass is the most relevant one 
leading to about a $4.5$~MeV effect on the prediction of $M_W$ for a 
top pole mass uncertainty of
$\delta m_t=0.75$~GeV. 
%The uncertainty has been reduced by additional LHC measurements but a
%slight tension between LHC and Tevatron oberservations
%exist~\cite{PDG:2018}. Further issues about relating the measured quantity
%to the actual mass parameter of the theory still remain.
While the experimental value given in eq.~\eqref{eq:mtexp} has a smaller
error, as discussed above it does not take into account the
systematic uncertainty arising from 
relating the measured mass parameter to a theoretically well-defined
quantity that can be used as input for the prtediction of $\MW$. 
The uncertainties on the hadronic contribution to
$\Delta \alpha$ and on $M_Z$ contribute up to 2.5~MeV each to the
uncertainty of the $\MW$ prediction.
The experimental error on the 
%Higgs boson mass $M_h$ and 
muon decay constant $G_F$ is negligible compared to the other sources. 
The same is true in the SM prediction for the experimental error on the
Higgs boson mass. 

%On the other hand, a theoretical uncertainty arises from missing 
%higher order effects.
The theoretical uncertainties from unknown higher-order contributions have
been estimated in ref.~\cite{Awramik:2003rn}
to be at the level of about 4~MeV in the SM for 
a light Higgs boson ($M_h^{\text{SM}}<300$~GeV).%
\footnote{In ref.~\cite{Degrassi:2014sxa} an
estimate of 6~MeV was given for the uncertainty by comparing the on-shell
prediction with a mixed-scheme calculation
done at the same perturbative order.} Uncertainties from unknown
higher-order SUSY loop contributions have been estimated for the MSSM and
the NMSSM in refs.~\cite{Heinemeyer:2006px,Stal:2015zca}. Since SUSY loop
effects decouple for heavy SUSY scales, also the uncertainty associated with
unknown higher-order SUSY contributions shrinks with higher SUSY masses,
while those uncertainties can be substantial for relatively light SUSY
states.
%Effects from SUSY are decoupling and the uncertainty shrinks with higher SUSY
%masses but they are relevant if lighter SUSY states are present.
At the two-loop level, corrections from gluinos, stops and sbottoms give the
largest effect in the MSSM. As explained above, contributions of this kind 
are not included in our MRSSM prediction because the
Dirac nature of the gluino in the MRSSM would require a dedicated
calculation of MRSSM two-loop contributions. 
Nevertheless, the contributions in the MSSM and MRSSM can be expected to be
of similar size, and we estimate an uncertainty of at most 
5~MeV for $m_{\text{SUSY}}>1$~TeV from the $\mathcal{O}(\alpha\alpha_s)$ 
corrections. As usual, the dependence of the MRSSM prediction on unknown
values of BSM parameters is not treated as a theoretical uncertainty, but
this dependence rather indicates the level of sensitivity for constraining
those parameters by confronting the $\MW$ prediction in the MRSSM with the
experimental result for this high-precision observable.

\section{Numerical results}
\label{sec:results}

In the following, we show how the prediction for $M_W$ in the MRSSM depends
on the general SUSY mass scale of the model. Then, we discuss how the
different SUSY sectors affect the prediction.
The SM input parameters~\cite{PDG:2018} are always set as follows
\begin{align*}
m_t &=173.00~\text{GeV}\,,\; \
\hat m^{\overline{\text{SM}}}_b(\hat m_b)=4.18~\text{GeV}\,,\;
M_Z=91.1887~\text{GeV}\,,\; \Gamma_Z=2.4952~\text{GeV}\,,\\
\Delta\alpha_{\text{lep}} &=0.031497686\,,\; 
\Delta\alpha^{(5)}_{\text{had}}=0.02761\,,
\alpha^{-1} = 137.035999139\,,\; 
\alpha^{\overline{\text{SM}}}_s(M_Z)=0.1181\,,\\
G_F &= 1.1663787\times 10^{-5}~\text{GeV}^{-2}\,,
\end{align*} 
where for $\Delta\alpha^{(5)}_{\text{had}}$ the result~\cite{Davier:2017zfy}
is used.

\subsection{General SUSY contributions and decoupling behaviour}
\label{sec:gen_results}

\begin{figure}
\includegraphics[width=0.5\textwidth]{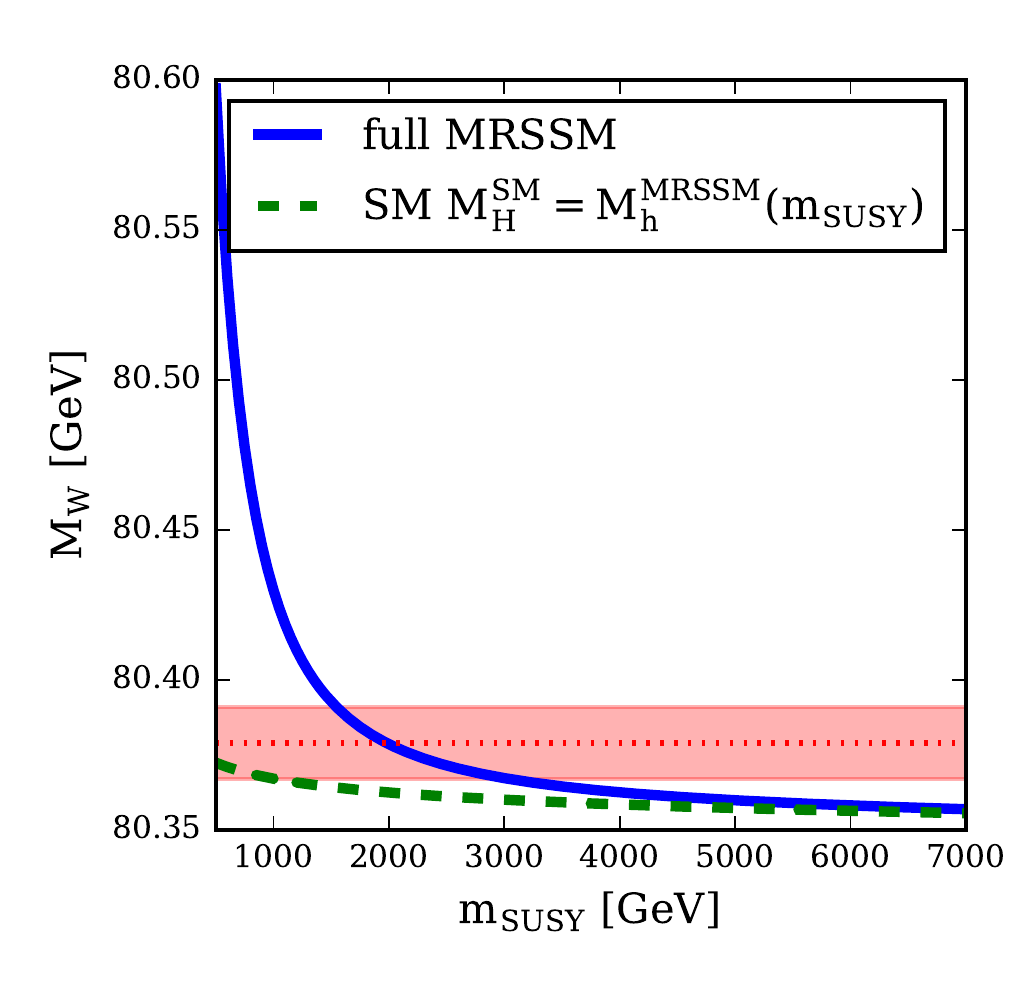}
\includegraphics[width=0.5\textwidth]{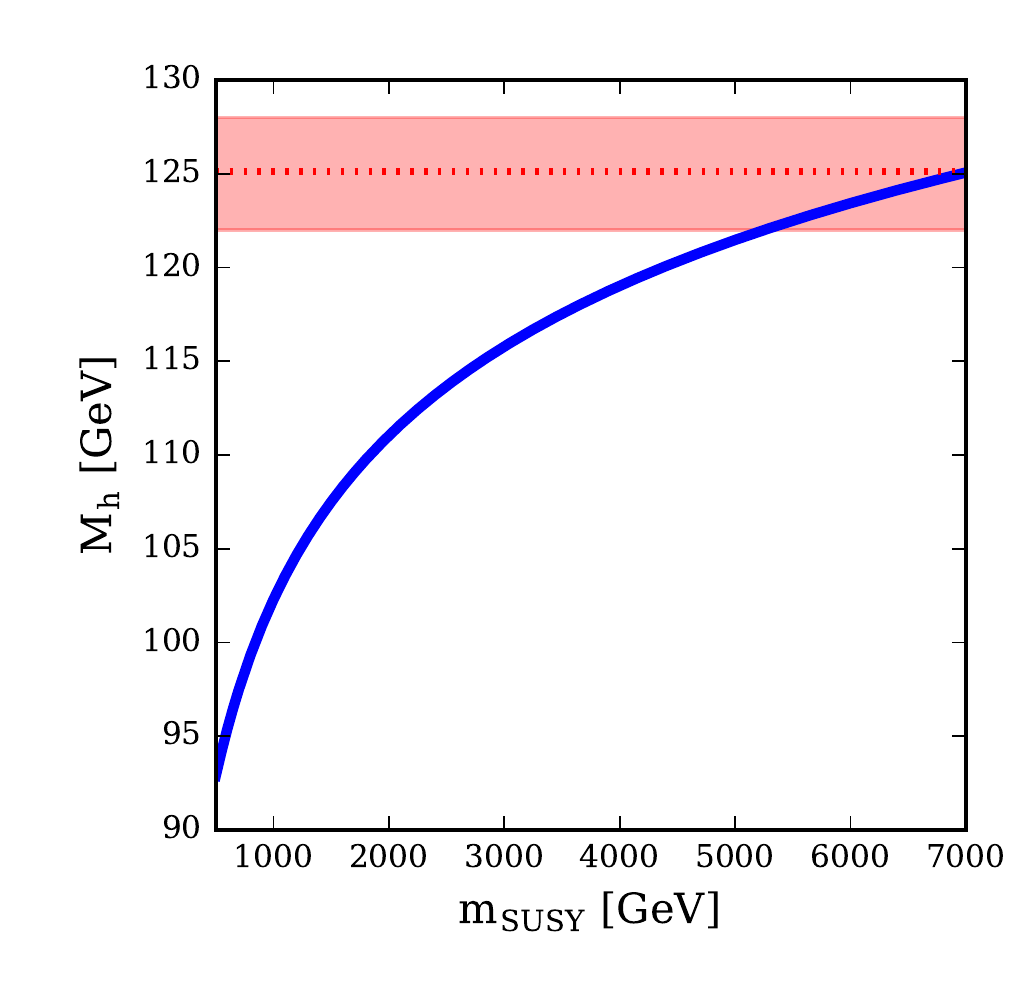}
\caption{Comparison of the $M_W$ ($M_h$) prediction as blue full line on the
left (right) depending on a common SUSY mass scale $m_{\text{SUSY}}$
(definition given in the text). 
The green dashed line on the left shows the prediction of the 
SM calculation for $M_W$ if
$M_h^{\text{SM}}=M_h^{\text{MRSSM}}(m_{\text{SUSY}})$ is used.
The red dotted line and the shaded area show on the left 
the experimental central value for $\MW$
and its 1~$\sigma$ uncertainty band according to
eq.~\eqref{eq:mw_value_exp_all},
while on the right the experimental value for the mass of the detected Higgs boson
is supplemented by a band of $\pm 3~\text{GeV}$ indicating a rough estimate 
of theoretical
uncertainties from unknown higher-order corrections. 
All dimensionless superpotential parameters are chosen as for BMP1 of
ref.~\cite{PD2}.
}
\label{fig:mw_msusy}
\end{figure}

We investigate the decoupling behaviour for $M_W$ by defining a general SUSY mass scale $m_{\text{SUSY}}$ as follows for the soft breaking parameters of eqs.~\eqref{eq:softmasses} and~\eqref{eq:mdirac}
as well as the $\mu_d$ and $\mu_u$ parameter of the superpotential:
\begin{align}
m^2_{R_u}=m^2_{R_d}=m_S^2=m_T^2=m_O^2=2\frac{B_\mu}{\sin
2\beta}&=m_{\text{SUSY}}^2\;, \nonumber\\
m^2_{\tilde q,L}=m^2_{\tilde l,L} =m^2_{\tilde e, R}=m^2_{\tilde
u,R}=m^2_{\tilde d,R}&=m_{\text{SUSY}}^2\cdot\mathbf{1} \;, \nonumber\\
M^D_B=M^D_W=M^D_O=\mu_d=\mu_u&=\frac{m_{\text{SUSY}}}{2}\;.
\label{eq:msusy}
\end{align}
The superpotential parameters are fixed to the value of the
benchmark point BMP1 given in ref.~\cite{PD2,Diss}, the dimensionless  couplings are set as $\Lambda_d=-1$, $\Lambda_u=-1.03$, $\lambda_d=1.0$ and
$\lambda_u=-0.8$. The ratio of the doublet vevs has been set to $\tan \beta=3$.

In figure~\ref{fig:mw_msusy} we show the dependence of $M_W$ and 
$M_h$ on the common SUSY mass scale $m_{\text{SUSY}}$ defined above.
The mass splitting between fermionic and bosonic mass parameters is required to 
achieve a prediction for the SM-like Higgs boson mass $M_h$ close to the 
experimental value measured at the LHC. The MRSSM prediction for $M_h$
is shown on the right-hand side of  fig.~\ref{fig:mw_msusy}. 
The prediction for $\MW$ in the MRSSM is compared in the left plot with the
SM prediction for $\MW$ where the mass of the Higgs boson in the SM is identified
with the corresponding MRSSM prediction for the mass of the SM-like Higgs
boson, 
$M_h^{\text{SM}}=M_h^{\text{MRSSM}}(m_{\text{SUSY}})$.
It can be seen that with a rising SUSY mass scale the SUSY contributions decouple,
and the prediction for $M_W$ approaches the SM limit for large SUSY masses. 
On the other hand, for small values of $m_{\text{SUSY}}$ the prediction for
$\MW$ shows a steep rise while the prediction for $M_h$ is significantly
lowered. As one can infer from a comparison of the two curves in the left
plot, the effect of shifting the value of the Higgs-boson mass in the SM-type
contributions accounts only for a a small fraction of the increase in $\MW$
for small $m_{\text{SUSY}}$, while the bulk of the effect is caused by
generic MRSSM  contributions. 

The comparison of the behaviour of the prediction for $\MW$ in 
figure~\ref{fig:mw_msusy} with the case of the MSSM and the NMSSM (see e.g.\ 
refs.~\cite{Heinemeyer:2006px,Stal:2015zca}) 
shows that in the MRSSM the decoupling to the SM
result occurs at higher values of the SUSY mass scale, while the
increase of $\MW$ for a decreasing SUSY scale is more pronounced than in the
MSSM and the NMSSM. The difference between the MRSSM prediction and the SM
prediction with $M_h^{\text{SM}}=M_h^{\text{MRSSM}}(m_{\text{SUSY}})$ still
amounts to about 10~MeV for $m_{\text{SUSY}}$ values of about 3~TeV in 
figure~\ref{fig:mw_msusy},
and the difference reduces to values below 1~MeV only for 
$m_{\text{SUSY}}\gsim 5~\text{TeV}$.
The different behaviour in the MRSSM as compared to the 
MSSM and the NMSSM~\cite{Heinemeyer:2006px,Stal:2015zca} 
%the prediction for $M_W$ approaches
%the SM value for a much lower SUSY mass scale.
%This can be explained by several factors.
is in particular related to the
enlarged matter content in the MRSSM, where the additional degrees of
freedom arise from the adjoint and the R-Higgs superfields, and to the fact
that the  $\Lambda/\lambda$ superpotential parameters contribute 
to $M_W$ in a similar way as the top Yukawa coupling.
In the considered scenario the $\Lambda_{u/d}$ couplings are of order one
and have a large effect on the prediction.
For low SUSY masses the triplet vev has a relevant influence on $M_W$
already at tree-level. The impact of the triplet vev
becomes negligible for larger SUSY mass scales as
$v_T$ goes to zero, see the discussion in section~\ref{sec:vt}.

\begin{figure}
\includegraphics[width=0.5\textwidth]{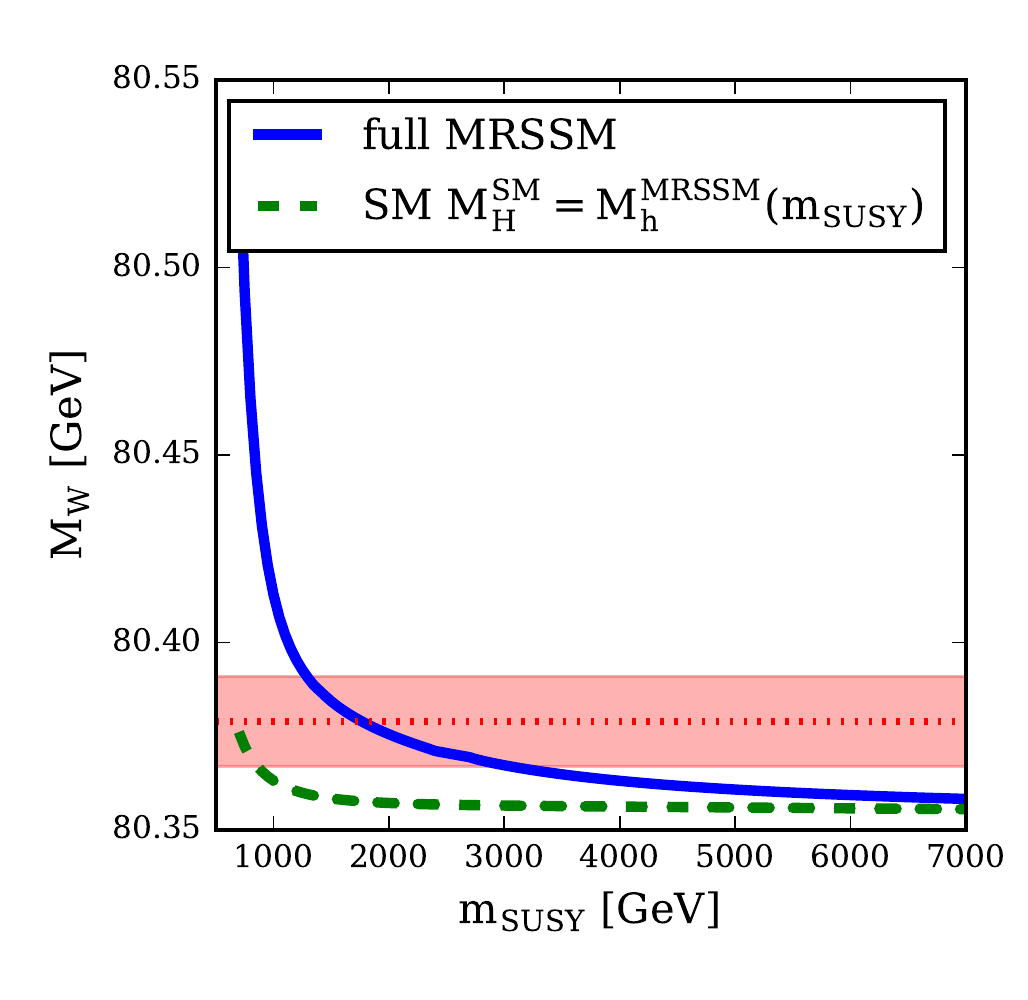}
\includegraphics[width=0.5\textwidth]{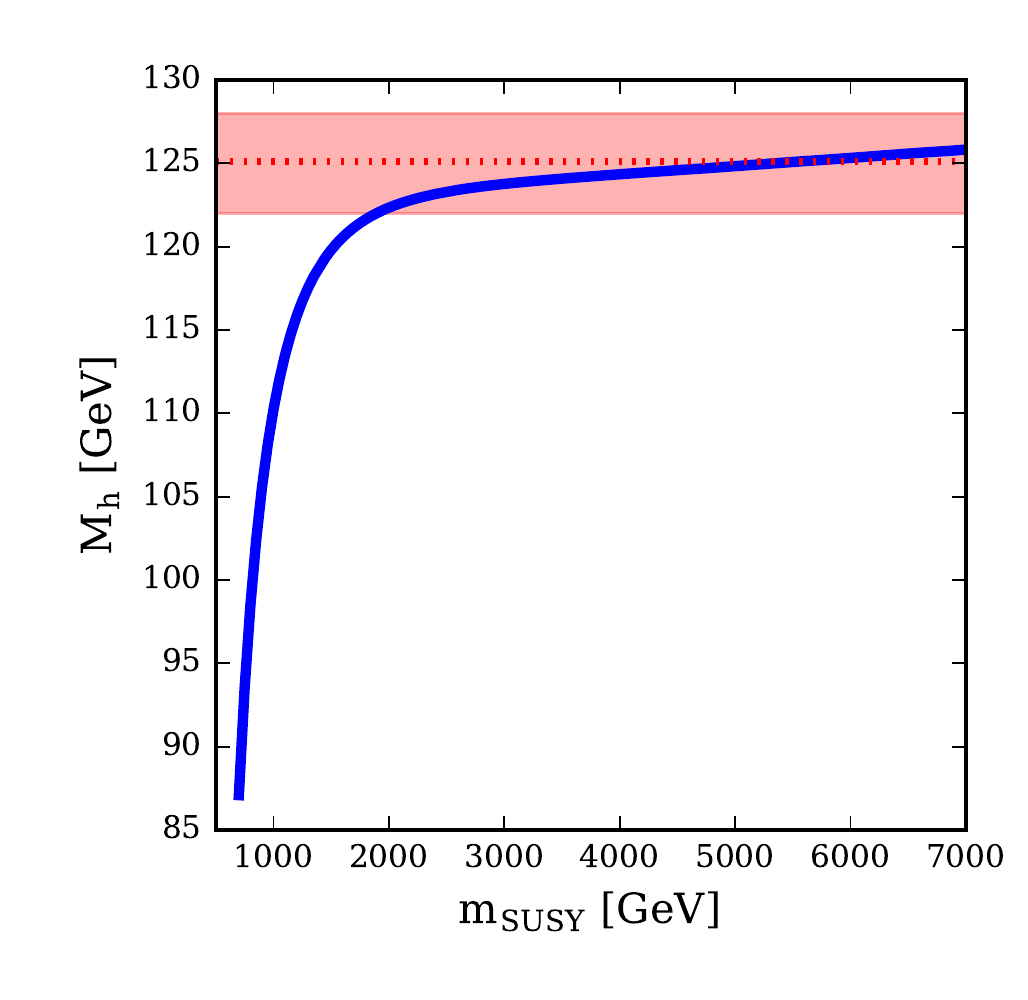}
\caption{
As figure~\ref{fig:mw_msusy} but with $\mu_d=\mu_u$ fixed to $400$~GeV.
}
\label{fig:mw_msusy_muconst}
\end{figure}

In the simplified scenario with a common mass scale 
$m_{\text{SUSY}}$ of figure~\ref{fig:mw_msusy} one can see that the range of 
$m_{\text{SUSY}}$ values yielding a prediction for $\MW$ within the $1\sigma$
band of the experimental central value does not coincide with the parameter
region where the mass of the SM-like Higgs boson is close to 125~GeV. The
prediction for $M_h$ in the MRSSM has a significant dependence on the
parameters $\mu_d$ and $\mu_u$. This can be seen in 
figure~\ref{fig:mw_msusy_muconst} where the same parameters as in 
figure~\ref{fig:mw_msusy} are used except that 
$\mu_d$ and $\mu_u$ are not scaled with 
$m_{\text{SUSY}}/2$ as in eq.~\eqref{eq:msusy} but kept fixed at 
$\mu_d=\mu_u = 400$~GeV. For fixed values of 
$\mu_d$ and $\mu_u$ the MRSSM prediction for $\MW$ still approaches the SM
prediction for large values of 
$m_{\text{SUSY}}$, 
but the numerical difference between the two predictions remains sizeable up
to even higher values of $m_{\text{SUSY}}$ than in 
figure~\ref{fig:mw_msusy}. On the other hand, the fixed value of 
$\mu_d=\mu_u = 400$~GeV brings the regions of 
$m_{\text{SUSY}}$ that are preferred by the $\MW$ and $M_h$ predictions into
better agreement with each other.

\subsection{Impact of different MRSSM contributions}

In the following we investigate how the different contributions in the MRSSM
affect $M_W$ separately. First we describe 
 the influence of the triplet vev on the prediction
for $M_W$. Then we describe how the different MRSSM sectors contribute to 
$M_W$ individually. 
Especially the extended Higgs and neutralino sectors are relevant in this
context as they differ
in the MRSSM from the MSSM and NMSSM and contain the effects
from the $\Lambda/\lambda$ couplings.
Several of the figures shown in the following sections contain plots of both
$M_W$ and $M_h$ as function of the parameters of interest. 
This is of interest as the Higgs boson mass in the MRSSM is very sensitive
to those parameters,
and as shown in figure~\ref{fig:mw_msusy} the variation of $M_h$ has an
impact on $M_W$ via
the SM-type contributions. 
In order to disentangle this contribution 
from the genuine MRSSM effects it is convenient to 
also show the dependence of $M_h$ on the relevant parameters.

The fixed parameters are set either as before, when $m_{\text{SUSY}}$ is 
varied, or we use updated values for BMP1 of ref.~\cite{PD2} giving rise to
a better agreement with the latest experimental value for $M_W$ given 
in~\eqref{eq:mw_value_exp_all}. The latter parameters are
\begin{align}
\tan\beta&=3,\; \Lambda_d=-1.2,\; \Lambda_u=-1.1,\; \lambda_d=1.0,\;
\lambda_u=-0.8,\notag\\ 
\mu_u&=\mu_d=500\text{ GeV},\; M_B^D=550\text{ GeV},\; M_W^D=600\text{ GeV},\;m_{R_d}=m_{R_u}=m_S=2\text{ TeV},\notag\\
 M_O^D&=1.5\text{ TeV},\; m_{\tilde l,L}=m_{\tilde e,R}=1\text{ TeV},m_{O}=m_{\tilde q,L;3}=m_{\tilde u,R;3}=m^2_{\tilde d,R;3}=1.5\text{ TeV},\;\notag\\
B_\mu&=(500\text{ GeV})^2,\; m_T=3\text{ TeV},\; m_{\tilde q,L;1,2}=m_{\tilde u,R;1,2}=m_{\tilde d,R;1,2}=2.5\text{ TeV}\;.
\label{eq:bm}
\end{align}

\subsubsection{Influence of the triplet vev}
\label{sec:vt_num}

\begin{figure}
\centering
\includegraphics[width=0.5\textwidth]{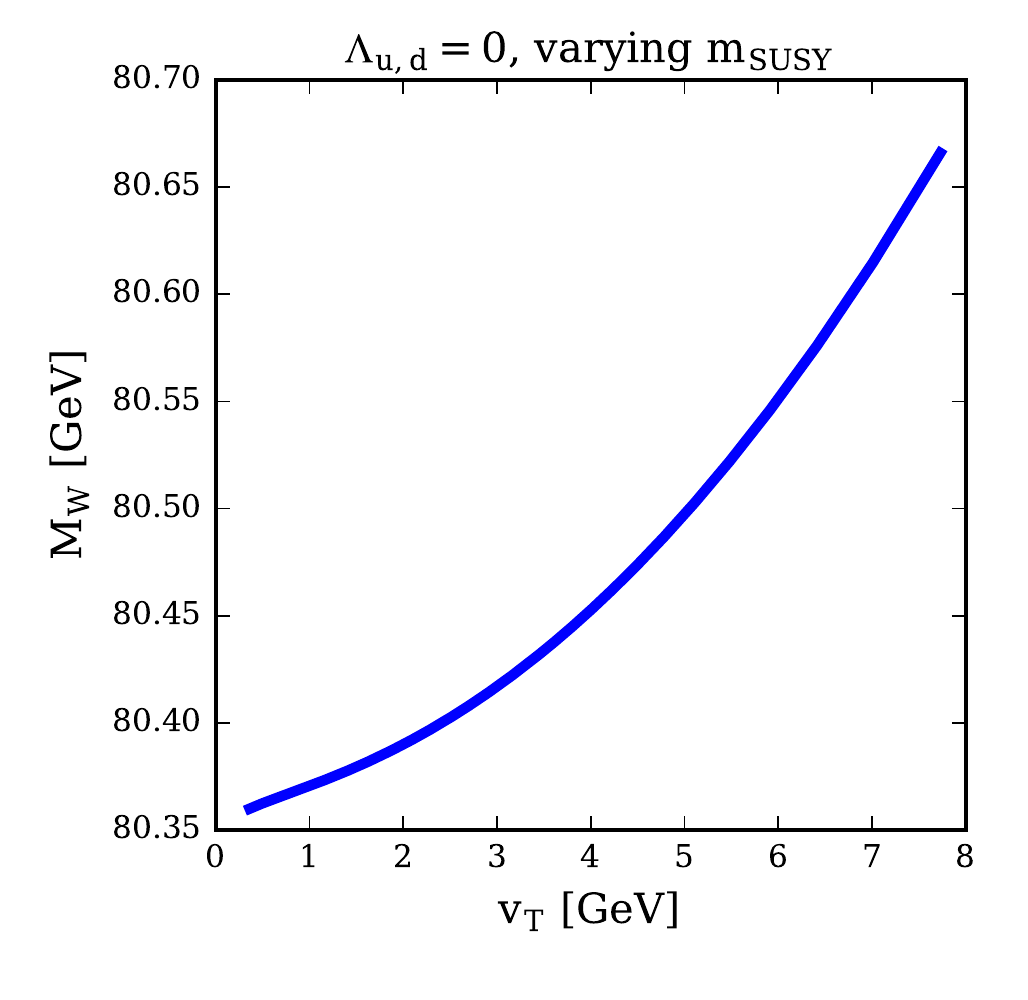}
\caption{
Dependence of $M_W$ on $v_T$ obtained from
varying the common SUSY mass scale as in figure~\ref{fig:mw_msusy}. 
}
\label{fig:vtmw}
\end{figure}

As the triplet vev affects the W boson mass already at the tree level by
breaking custodial symmetry, even a small value (compared to $M_W$) affects
the prediction at the same order as the size of the experimental uncertainty.
%We take the usual approach to highlight this where a limit on the triplet 
%vev is derived from the $\rho_0$ parameter defined in 
%ref.~\cite{Patrignani:2016xqp}.
%Then an limit for $v_T^2$ can be derived from the value derived by the global fit
%\begin{equation}
%\rho_0^{exp}=1.00037\pm0.00023\;,
%\end{equation}
%where $v_T$ contributes as, see eq. (10.62) of ref.~\cite{Patrignani:2016xqp},
%\begin{equation}
%\rho_0=1+\frac{4v_T^2}{v^2}\;.
%\end{equation}
%Taking the central value originating from this contribution while the uncertainty might account for further loop effects yields an upper bound of 
%$|v_T^{\text{tree}}|\lesssim 3$~GeV.
%Assuming that this is the only BSM influence applied to the prediction of
%$M_W$ the numerical shift is $\delta M_W^{(v_T)}\approx35\text{ MeV}$.

In figure~\ref{fig:vtmw} we show the interplay of $v_T$ and $M_W$ when the SUSY 
mass scale $m_{\text{SUSY}}$ is varied as in figure~\ref{fig:mw_msusy}
and all $\lambda_{d,u}/\Lambda_{d,u}$ are 
set equal to zero. One can see that in this case 
the $v_T$ tree level contribution is numerically large
for $v_T \gsim 1$~GeV. The potentially large impact of $v_T$ can be clearly
seen by the quadratic dependency exhibited in the figure which is in
accordance with eq.~\eqref{eq:masses}. For $|v_T| \gsim 3$~GeV the
prediction for $M_W$ grows above the experimentally allowed region. Therefore,
for phenomenological reasons the 
parameter region with $|v_T| \lsim 3$~GeV is preferred.
For $|v_T|\approx 3$~GeV, 
using the parameters of eq.~\eqref{eq:bm} and adjusting $m_T$ accordingly, 
the  associated
one-loop contribution (which is proportional to $v_T^2$) to $\Dhr$ from 
eq.~\eqref{eq:dr-start} is of similar size as the SM three- and four-loop
contributions 
%as the different contributions partially cancel each other 
leading to a shift of $\delta M_W^{(v_T),\text{Loop}}\approx 1\text{ MeV}$.
It should be noted that the higher-order contributions
also significantly depend on the SUSY mass spectrum.

For our numerical analysis in the following we choose 
the parameters of eq.~\eqref{eq:bm} as basis.
This yields
%The benchmark point chosen for this paper has 
\begin{equation}
v_T^{\DR}(Q=m_{\text{SUSY}})=-0.38~\text{GeV} 
\label{eq:vTvalue}
\end{equation}
as an input for our $M_W$ calculation.
%using the parameters of eq.~\eqref{eq:bm} and adjusting $m_T$ accordingly. 
This setting leads to a 
a tree-level (one-loop) shift of $\delta M_W^{(v_T)}=0.28$~MeV
($\delta M_W^{(v_T)}=-0.17$~MeV), respectively,
where again it should be noted that the higher-order contributions
are significantly affected by the parameters of the SUSY mass spectrum.

\subsubsection{Influence of the $\Lambda$ couplings}

\begin{figure}
\includegraphics[width=0.5\textwidth]{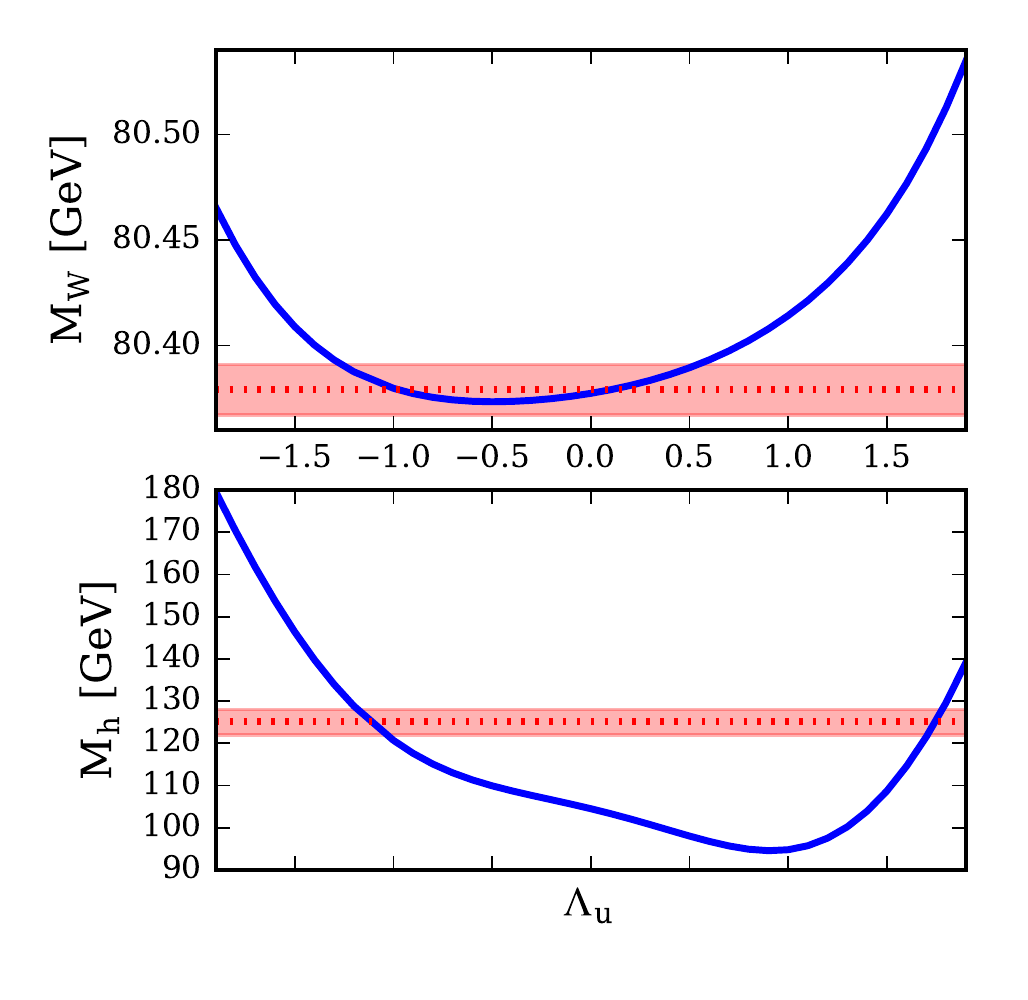}
\includegraphics[width=0.5\textwidth]{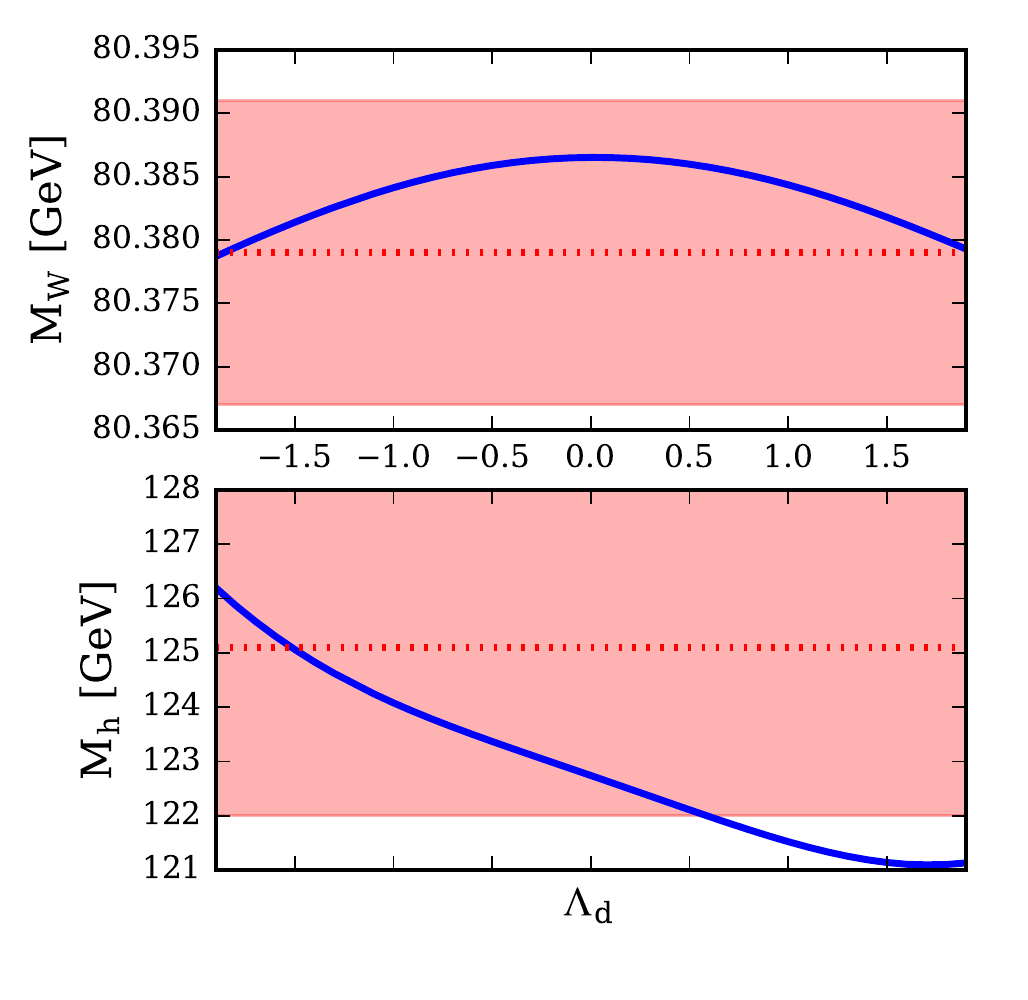}
\caption{
Dependence of $M_W$ and $M_h$ on $\Lambda_d$ and $\Lambda_u$.
The value for $v_T$ is fixed according to eq.~\eqref{eq:vTvalue}, and
the other model parameters are chosen as 
%for BMP1 of ref.~\cite{PD2}.
in eq.~\eqref{eq:bm}.
The red bands mark the experimental (theoretical) uncertainty region
for $M_W$ ($M_h^{\text{SM-like}}$)
as in figure~\ref{fig:mw_msusy}.}
\label{fig:lambdas}
\end{figure}

The dependence of $M_W$ on the superpotential 
parameters $\Lambda_d$ and $\Lambda_u$
is shown in figure~\ref{fig:lambdas}.
The effects are stronger for $\Lambda_u$ than for $\Lambda_d$
because of $\tan\beta>1$. In the extreme case of 
$\Lambda_u=2$ 
a shift of more than $\delta M_W^{(\Lambda_u)}=130$~MeV 
is possible compared to the minimal value at
$\Lambda_u=-0.5$. 
In the phenomenologically interesting region where the MRSSM Higgs boson mass
prediction is around the experimental observation of about $125$~GeV with
$\Lambda_u=-1.1$ 
the shift in $M_W$ compared to $\Lambda_u=0$ is 
$\delta M_W^{(\Lambda_u)}\approx10$~MeV.
The actual effect from the $\Lambda_u$ variation on $M_W$ 
for large $|\Lambda_u|$ 
is even stronger than the variation displayed in 
figure~\ref{fig:lambdas} since 
large $|\Lambda_u|$ also drives the SM Higgs mass prediction to values
far above the
experimental observation,
as shown in the lower left plot of figure~\ref{fig:lambdas},
which gives rise to a downward shift
from the SM-type part of the contributions to $M_W$.
As shown on the right-hand side of figure~\ref{fig:lambdas}, 
an enhancement of the magnitude of $\Lambda_d$ from zero to unity leads to a
reduction of $\delta M_W^{(\Lambda_d)}=-5$~MeV, while the prediction for
$M_h$ is lowered by about 1~GeV.

The behaviour of $M_W$ with regard to the two 
parameters $\Lambda_d$ and $\Lambda_u$
can be understood
from their influence on the electroweak 
precision parameters $S$, $T$ and $U$ which contribute to $M_W$. 
For this, we follow the lines of refs.~\cite{PD1,Diss}. 
There, several limits of the contributions to $S$, $T$ and $U$ were studied analytically. The $T$ 
parameter has been identified as the one with the biggest impact on $M_W$,
where the contributions from the neutralino and chargino sector dominate.
Up to a prefactor the definition of the $T$ parameter corresponds
to the one of $\Delta\rho$.
In general, the $\Delta\rho$ parameter depends on the fourth power of $\Lambda$
leading to significant effects for magnitudes close to or above unity,
as it is visible in figure~\ref{fig:lambdas}.

It is of interest to understand the interplay of $\Lambda_d$ and $\Lambda_u$ 
concerning the $M_W$ prediction. In the previous works of 
refs.~\cite{PD1,Diss}, all analytical expressions 
in the study of several model limits have 
been derived setting $v_d=0$, turning off all $\Lambda_d$ contributions.
As we want to investigate these contributions we discuss a different model limit
in the following.
We take the gaugeless limit ($g_1=g_2=0$) and set $\mu_d=\mu_u=M^D_W$ and
$\lambda_u=\lambda_d=0$.
Then, we find contributions from the electroweakino sector to the 
$\Delta\rho$ parameter as
\begin{equation}
\Delta \rho_0^{\Lambda} = \frac{\alpha}{16\pi M_W^2\smixx}\frac{13\left(\Lambda_u^2 v_u^2-\Lambda_d^2 v_d^2\right)^2}{96 (M_W^D)^2}.
\label{eq:tpar-lam}
\end{equation}
The relative sign between the two couplings and the fact that $\tan\beta>1$ 
for the scenario considered is the reason why an increase of $\Lambda_d^2$ actually
decreases the prediction for $M_W$ when $\Lambda_d$ and $\Lambda_u$ are of similar magnitude. The
$\Lambda_u$ contribution dominates for most of the parameter space,
and the $\Lambda_d^2$ term leads to a reduction of the contribution in this
case.
The situation is reversed for $\tan\beta<1$.

\begin{figure}[htb]
\includegraphics[width=0.5\textwidth]{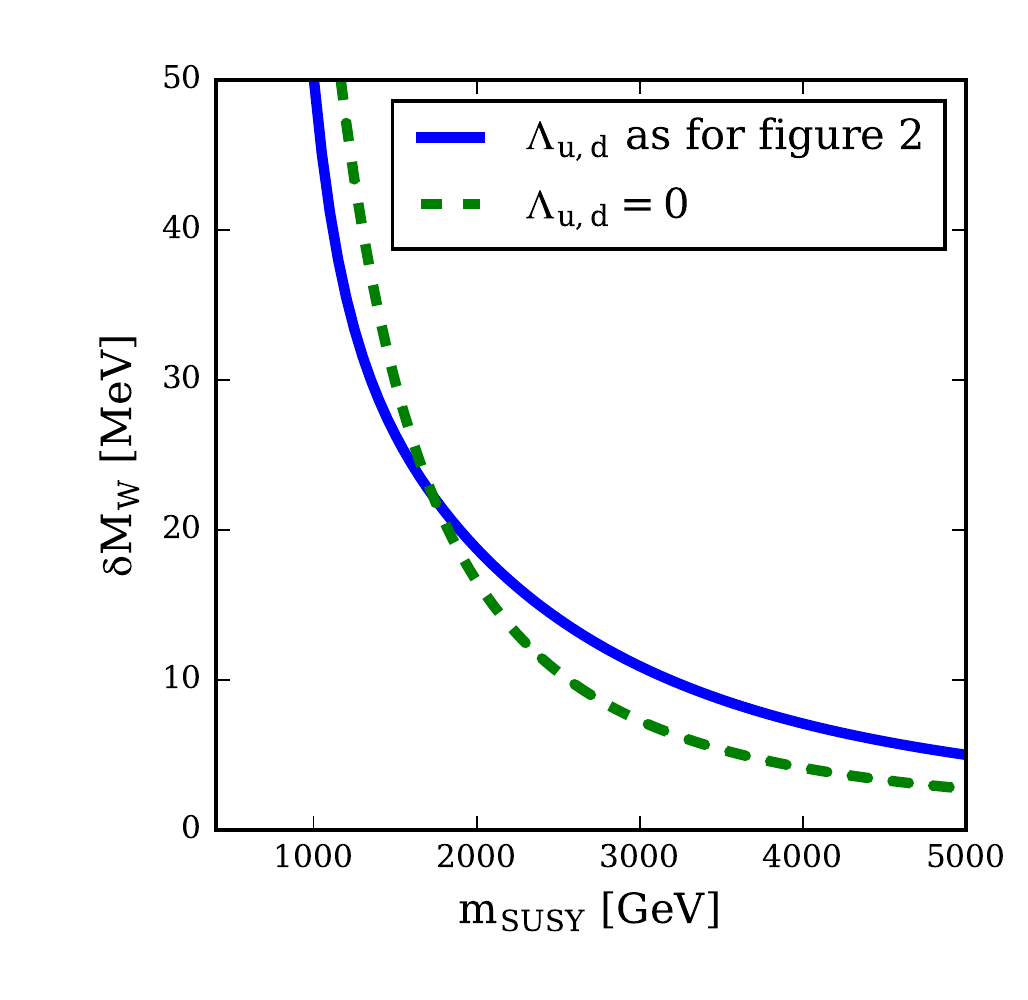}
\includegraphics[width=0.5\textwidth]{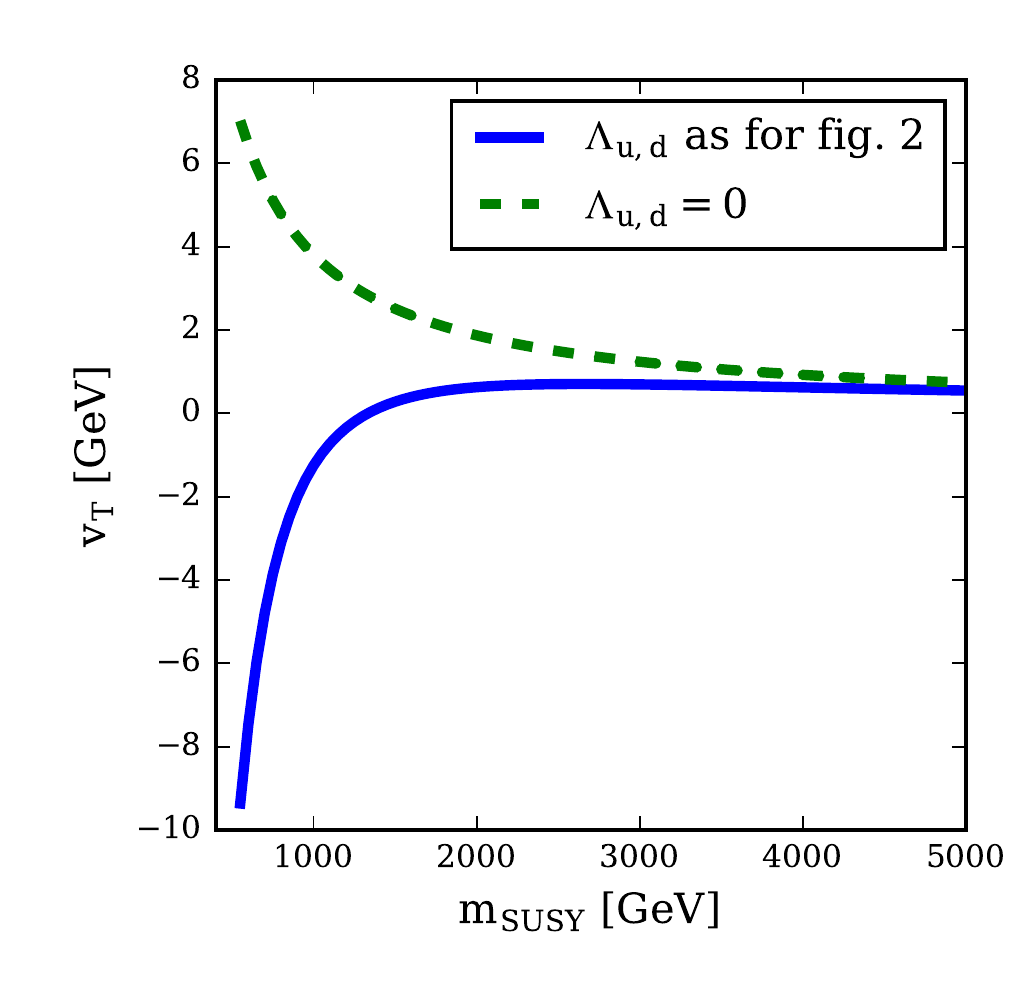}
\caption{
The plot on the left depicts the shift in the prediction of $M_W$ with
respect to the SM prediction according to
eq.~\eqref{eq:dmw}.
On the right the corresponding value of $v_T$ is plotted. Two different scenarios
are considered where the parameters $\Lambda_{d,u}$ are fixed to either
the values $\Lambda_d=-1$, $\Lambda_u=-1.03$ or to zero, 
while $\mu_d=\mu_u=500$~GeV, 
and the values of all other mass parameters are
given in eq.~\eqref{eq:msusy}.
}
\label{fig:lambdavt}
\end{figure}

In figure~\ref{fig:lambdavt} we show on the left
the quantity
\begin{equation}
\delta M_W = M_W^{\text{MRSSM}}-M^W_{\text{SM}}
(M_H^{\text{SM}}=M_h^{\text{MRSSM}}(m_{\text{SUSY}}))\,,
\label{eq:dmw}
\end{equation} 
where as above $M_h^{\text{MRSSM}}$ is the SM-like Higgs boson mass of the MRSSM.
On the right the prediction for $v_T$ is depicted. We compare the scenario
studied above where 
the magnitudes of $\Lambda_{d,u}$ are of order unity 
with a scenario where $\Lambda_d$ and $\Lambda_u$ are both fixed
to zero. Note that it is necessary to fix the values of $\mu_d$ and $\mu_u$
to the ones of the benchmark point of eq.~\eqref{eq:bm}, while all other
dimensionful
SUSY parameter are scaled with $m_{\text{SUSY}}$ as specified in 
eq.~\eqref{eq:msusy}. 
The parameters $\lambda_d$ and $\lambda_u$ are also set as in eq.~\eqref{eq:bm}.
This setting limits the size of
$v_T$ for $m_{\text{SUSY}}$ below 1~TeV.
In both scenarios the difference in the prediction for $M_W$ between
the MRSSM and the 
SM decreases from more than $50$~MeV at $m_{\text{SUSY}}\approx 1$~TeV
to below 5~MeV for $m_{\text{SUSY}}$ values in the multi-TeV range. 
Both lines show a very similar shape while the 
main contributions
to $\delta M_W$ are different in the two scenarios. 

In the scenario with $\Lambda_{d,u}$ of about $-1$ the magnitude of the triplet vev
drops to below 1~GeV for $m_{\text{SUSY}}>1$~TeV as there is a partial
cancellation in the numerator of eq.~\eqref{eq:vt}
between the terms of $\Lambda_{u/d} \mu_{u/d}$  and $g_2 M_W^D$, which are
of similar magnitude. 
Therefore, $v_T$ drops below 1~GeV already for relatively
low  $m_{\text{SUSY}}$.
It goes to zero for larger $m_{\text{SUSY}}$ but with a smaller rate as the cancellation in  the numerator of eq.~\eqref{eq:vt} is not perfect.
As $v_T$ is small above $m_{\text{SUSY}}=1$~TeV, its tree-level contribution 
does not substantially impact the prediction for $M_W$ in this parameter
region. Then, loop corrections
from $\Lambda_{d,u}$ like eq.~\eqref{eq:tpar-lam} are the most relevant ones.

In the scenario where
$\Lambda_{d,u}$ are fixed to zero their loop contributions vanish.
On the other hand, there is no significant cancellation in the numerator of
eq.~\eqref{eq:vt} for $\tan\beta\gg1$. 
Then, $v_T$ drops below 1~GeV only above
$m_{\text{SUSY}}=3.5$~TeV, and its tree-level contribution to $M_W$
is relevant for a larger mass range.

\subsubsection{Higgs sector contributions}

\begin{figure}
\includegraphics[width=0.5\textwidth]{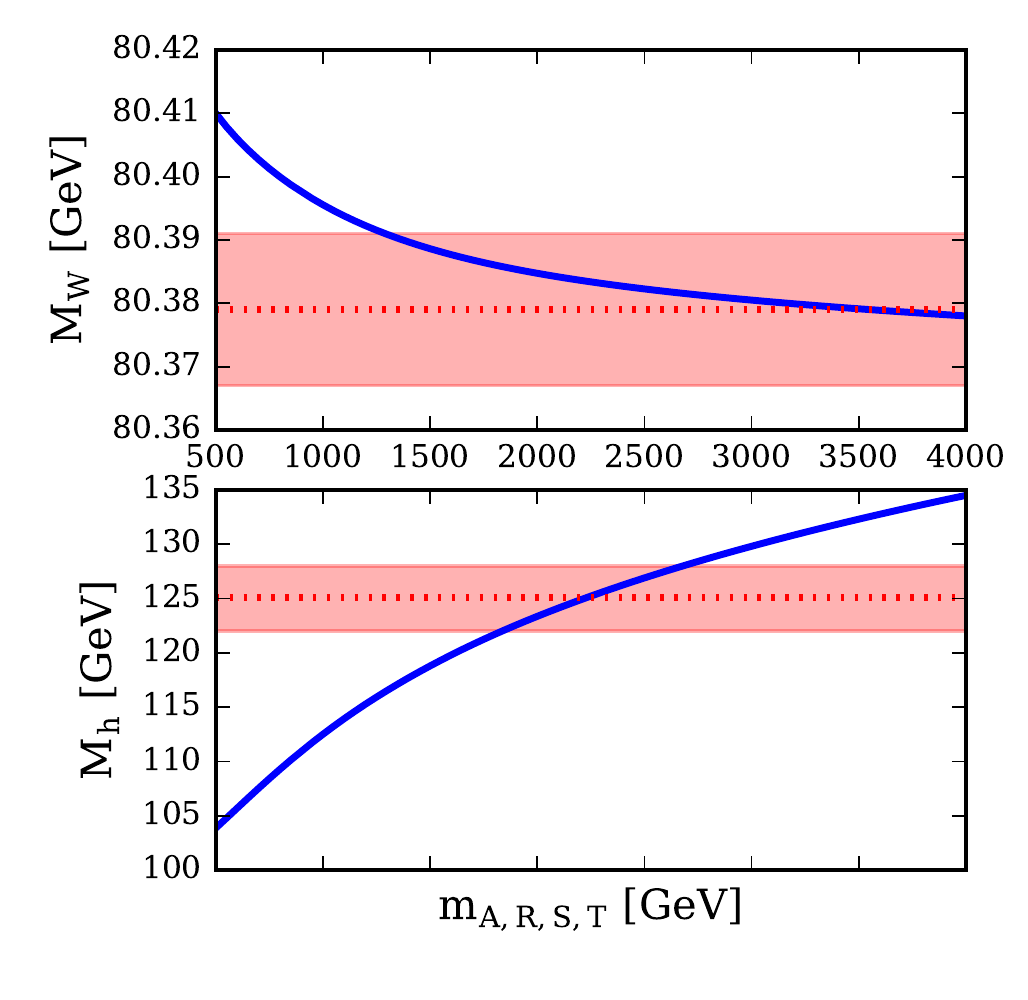}
\includegraphics[width=0.5\textwidth]{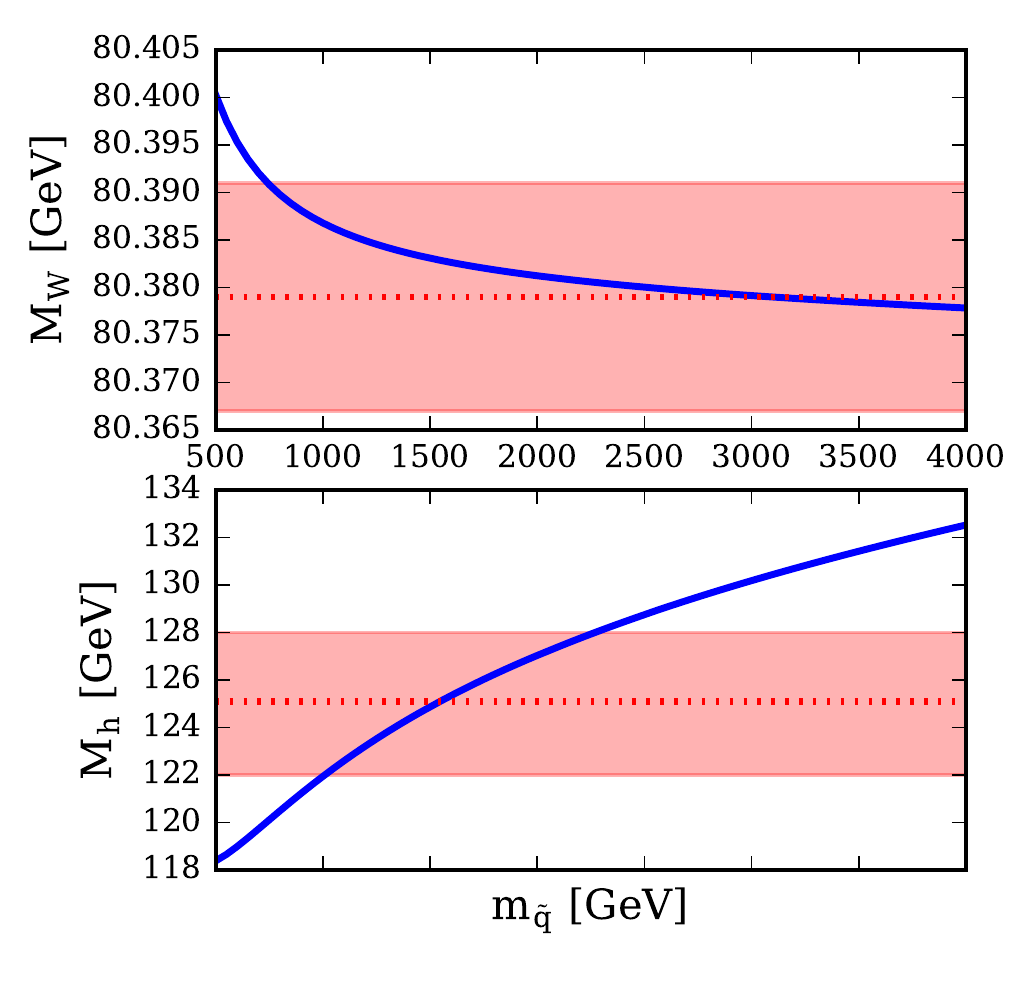}
\caption{
Dependence of $M_W$ and $M_h$ on a common Higgs sector mass (left) and a
common squark mass (right). The other parameters are chosen as in
figure~\ref{fig:lambdas}.}
\label{fig:softmasses}
\end{figure}

In the following, we investigate the influence of the extended Higgs sector 
of the MRSSM on the W boson mass.
The contributions from the Higgs sector include the ones from the two MSSM-like Higgs doublets, the singlet and
triplet states, which all mix with each other, as well as the two R-Higgs doublets.
We vary all the relevant soft breaking masses $m_S^2$, $m_T^2$, $m_{R_{u/d}}^2$ as well as the
MSSM-like CP-odd Higgs mass parameter $m_A^2=2 B_\mu/\sin 2\beta$ simultaneously to show the dependence of the W boson mass on them.
On the left side of figure~\ref{fig:softmasses}  the dependences of $M_W$ and $M_h$ on these parameters are shown. As a comparison we show the
dependence on the squark masses on the right, here all the soft-breaking
squark mass parameters are varied simultaneously.

Altogether, varying the common Higgs sector mass from 500~GeV to 4~TeV
leads to an increase of the Higgs boson mass by almost 30 GeV and a
simultaneous decrease of $M_W$ by about 30~MeV.
%, where already in the range from 500 to 1500~GeV a drop by about 20~MeV is observed.
The variation of the common squark mass from 500~GeV to 4~TeV on the other 
hand increases the Higgs boson mass by about 14~GeV while $M_W$ drops by
about 20~MeV.
%where a 15~MeV difference occurs in the range from 500 to 1500~GeV.
For both plots decoupling-like behaviour of the relevant SUSY contributions
to $\MW$ can be seen which is similar to the one in
figure~\ref{fig:mw_msusy}. The residual slope for high SUSY masses
originates purely from
the Higgs mass dependence of the SM-like contributions to $\Dhr$.

It has been noted before~\cite{PD1} that ${\cal O}(1)$
$\lambda/\Lambda$ parameters have effects comparable to the ones of the top Yukawa coupling as
both, the Yukawa and $\lambda/\Lambda$ couplings, originate from similar
terms of the superpotential. The squark contributions to $M_W$ stem mainly
from top-Yukawa effects with a suppression factor originating from 
the squark masses. 
The Higgs sector effects on the other hand
are driven by the $\lambda/\Lambda$ parameters being of order one and are suppressed
by the soft-breaking Higgs mass terms.
Therefore, 
the similar behaviour arising from both sectors for $M_W$ and $M_h$ as shown in figure~\ref{fig:softmasses} 
is expected. 
The effect of the Higgs sector is quantitatively larger when varying the involved 
common mass parameters in a similar range as more degrees of freedom
contribute in the Higgs sector even taking into account the colour factor
for the squark contributions. A rising $m_T$ leads to a suppression
of the triplet vev and its contribution to $M_W$ via the tadpole
relation~\eqref{eq:vt}.

\subsubsection{Neutralino contributions}

\begin{figure}
\includegraphics[width=0.5\textwidth]{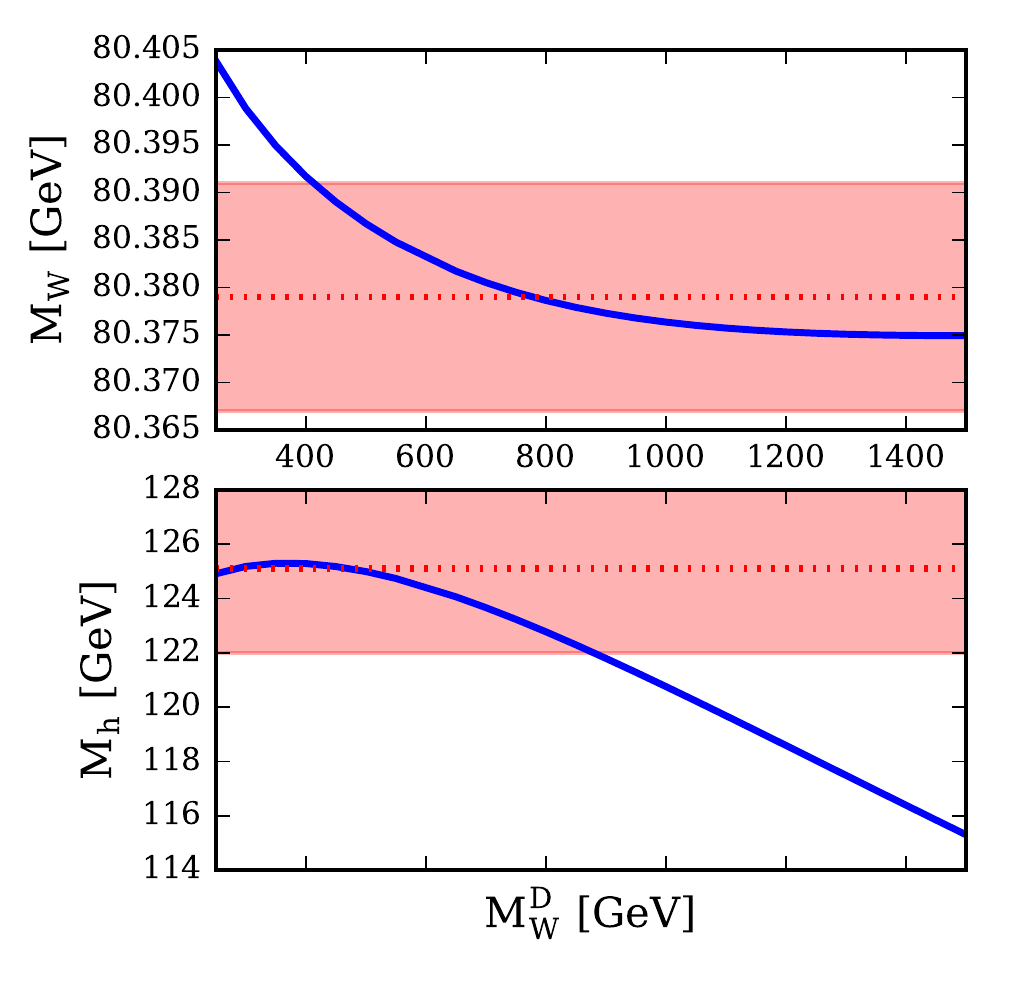}
\includegraphics[width=0.5\textwidth]{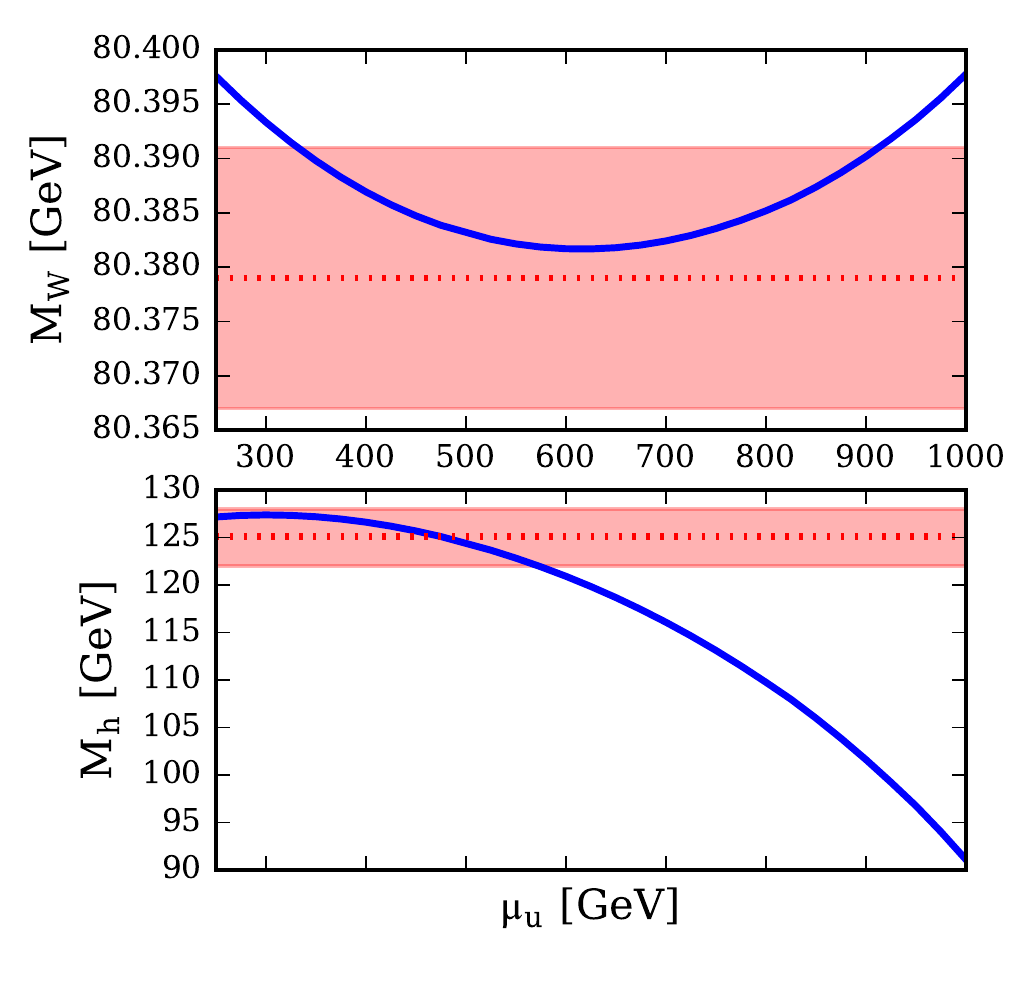}
\caption{
Dependence of $M_W$ and $M_h$ on $M^D_W$ (left) and $\mu_u$ (right).
The other parameters are chosen as in figure~\ref{fig:lambdas}.}
\label{fig:neutralinosec}
\end{figure}

Contributions of charginos and neutralinos to $M_W$ are very relevant in
the MRSSM as the $\Lambda$ superpotential parameters affect their mixing
and lead to novel contributions.
Additionally, the dimensionful parameters influence these corrections
directly and indirectly, which is discussed in the following.

Figure~\ref{fig:neutralinosec} (left) shows that
the effect of the mass parameter $M^D_W$ 
is similar to the
ones of the scalar soft breaking masses discussed before, i.e.\ the
predicted value for $M_W$ decreases for rising $M^D_W$.
As already noted in the discussion of 
figure~\ref{fig:mw_msusy} and figure~\ref{fig:mw_msusy_muconst}, the
parameter $\mu_u$ has a large impact on the 
prediction for $M_h$ in the MRSSM.
In the plots on the right-hand side of
figure~\ref{fig:neutralinosec}
the variation of $\mu_u$ is only shown in the range of $250$ to $1000$~GeV
as for higher values of $\mu_u$ 
the tree-level Higgs boson mass becomes tachyonic.
This is caused by the effect of $\mu_u$ on
the singlet--doublet mixing of the scalar Higgs boson mass matrix  in the form 
$\mathcal{M}_{13}^{\text{CP-even}}=-v_u(\sqrt{2}\lambda_u \muu{-}+g_1
M^D_B)$. Because the SM-like Higgs boson is the lightest 
electroweak scalar in the considered scenario, the enhanced mixing leads to
a strong
reduction of the Higgs boson mass and the appearance of a tachyonic state
for large $\mu_u$.
The Higgs boson mass also decreases with rising $M^D_W$ as the loop
contributions to $M_h$ depend on $\log (m_T/M^D_W)$. As the Dirac mass is
smaller than the scalar mass in the considered range the Higgs boson mass
decreases with increasing $M^D_W$.

Varying $M^D_W$ and $\mu_u$ 
in the mass range from 250 to 600 GeV yields a downward shift in the $M_W$ 
prediction of
$\delta M_W^{(M^D_W)}=21$~MeV for $M^D_W$ and $\delta M_W^{(\mu_u)}=16$~MeV 
for $\mu_u$.
The downward shifts are mainly driven
by contributions from 
$\Delta \rho_0^{\Lambda}$,
see eq.~\eqref{eq:tpar-lam}.
For the mass range above about $600$~GeV the prediction for the mass of the
SM-like Higgs boson decreases for both parameters,
as described above.
In this parameter region
the decrease in the Higgs boson mass leads to additional 
contributions shifting $\MW$ upwards. Those contributions from the SM-like
Higgs boson partially cancel the SUSY corrections, with a net effect in the 
$M_W$ prediction 
of a less steep decrease with rising $M^D_W$ and of an increase with 
rising $\mu_u$.

\subsection{Scan over MRSSM parameters}

\begin{figure}
\centering
\includegraphics[width=0.7\textwidth]{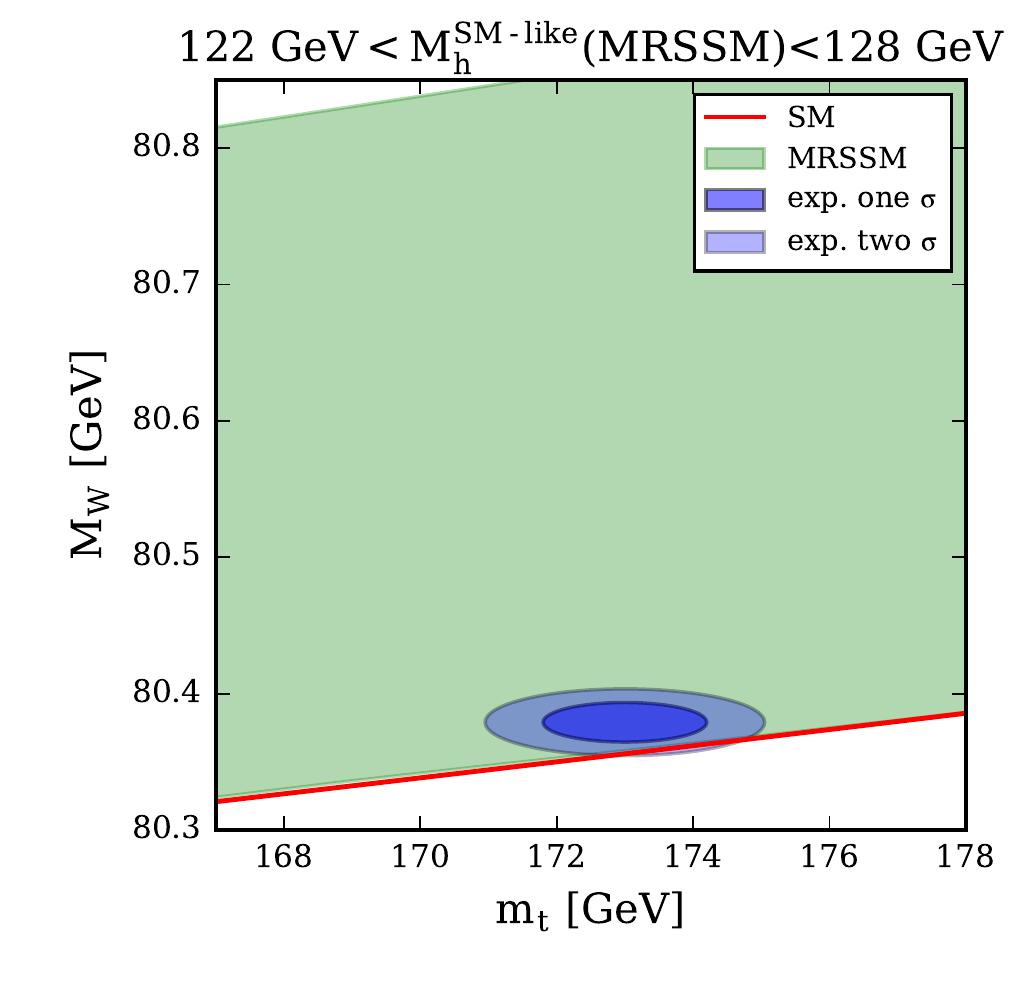}
\caption{
Prediction for the W boson mass 
in the MRSSM and the SM 
versus the input value of the top quark pole mass 
in comparison with the
experimental results for $\MW$, see
eq.~\eqref{eq:mw_value_exp_all},
and $m_t$, see
eq.~\eqref{eq:mtexp}.
The green band of MRSSM predictions arises from the scan over MRSSM
parameters specified in eq.~\eqref{eq:scan}, where the condition 
$122~\text{GeV} < M_h < 128~\text{GeV}$ has been applied for the
prediction of the SM-like Higgs boson mass. 
The narrow red band, which is shown on top of the green band, indicates the
SM prediction where
the experimental measurement is used as input for the Higgs boson mass.
For the experimental measurements of $\MW$ and $m_t$ the two-dimensional
regions allowed at the 1~$\sigma$ and the 2~$\sigma$ level are indicated.
}
\label{fig:mtmw}
\end{figure}

In figure~\ref{fig:mtmw} we show the predictions for $\MW$ 
in the MRSSM resulting from a scan over MRSSM parameters
as a green band
in relation to the input value for the top quark pole mass. 
We have arrived at this band by scanning over several parameter combinations,
\begin{align}
-1.5&<-\lambda_d=\Lambda_d=\lambda_u=\Lambda_u<1.5\;, \nonumber \\
250\text{ GeV}&<\mu_d=\mu_u=M^D_B=M^D_W<3000\text{ GeV}\;, \nonumber \\
250\text{ GeV}&<m_S=m_T=m_{R_{d}}=m_{R_u}=m_A<3000\text{ GeV}\;, \nonumber \\
250\text{ GeV}&<m_{\tilde q,L}=m_{\tilde l,L}=m_{\tilde d,R}=m_{\tilde
u,R}=m_{\tilde e,R}<3000\text{ GeV}\;,
\label{eq:scan}
\end{align}
where the prediction for the mass of the SM-like Higgs boson has been
required to agree with the measured value within the theoretical
uncertainties. As condition for $M_h$ the mass range 
$122~\text{GeV} < M_h < 128~\text{GeV}$ has been adopted. 
The SM prediction where the measured Higgs boson mass within
the experimental uncertainties 
has been used as input is shown in red.
The blue ellipses mark the measurements of the
top quark and the W boson mass including their two-dimensional 1~$\sigma$
and 2~$\sigma$ experimental uncertainty regions. 
It should be noted that the systematic uncertainty in relating the
measured value of the top quark mass to the top quark pole mass (see the
discussion above) is not
accounted for in figure~\ref{fig:mtmw}. A proper inclusion of this
uncertainty would widen the displayed ellipses along the $m_t$ axis. 

The comparison of the MRSSM and SM predictions for $\MW$ with the
experimental results for $\MW$ and $m_t$ in 
figure~\ref{fig:mtmw} shows on the one hand that MRSSM parameter regions
giving rise to a large upward shift in $\MW$ as compared to the SM case are
disfavoured by the measured value of $\MW$, in accordance with the results
of figures~\ref{fig:mw_msusy}--\ref{fig:neutralinosec}. On the other hand, 
figure~\ref{fig:mtmw} indicates a slight preference for a non-zero SUSY 
contribution to $\MW$, 
see also the discussion for the MSSM and the NMSSM 
in refs.~\cite{Heinemeyer:2006px,Heinemeyer:2013dia}. The band of the
SM prediction barely intersects with the 2~$\sigma$ ellipse of the 
experimental results in figure~\ref{fig:mtmw}, 
and a non-zero SUSY contribution giving rise to a moderate
upward shift in $\MW$ is required 
in order to reach compatibility with the experimental
result at the 1~$\sigma$ level.

\subsection{Comparison to other calculation methods}

\begin{figure}
\begin{center}
\includegraphics[width=0.45\textwidth]{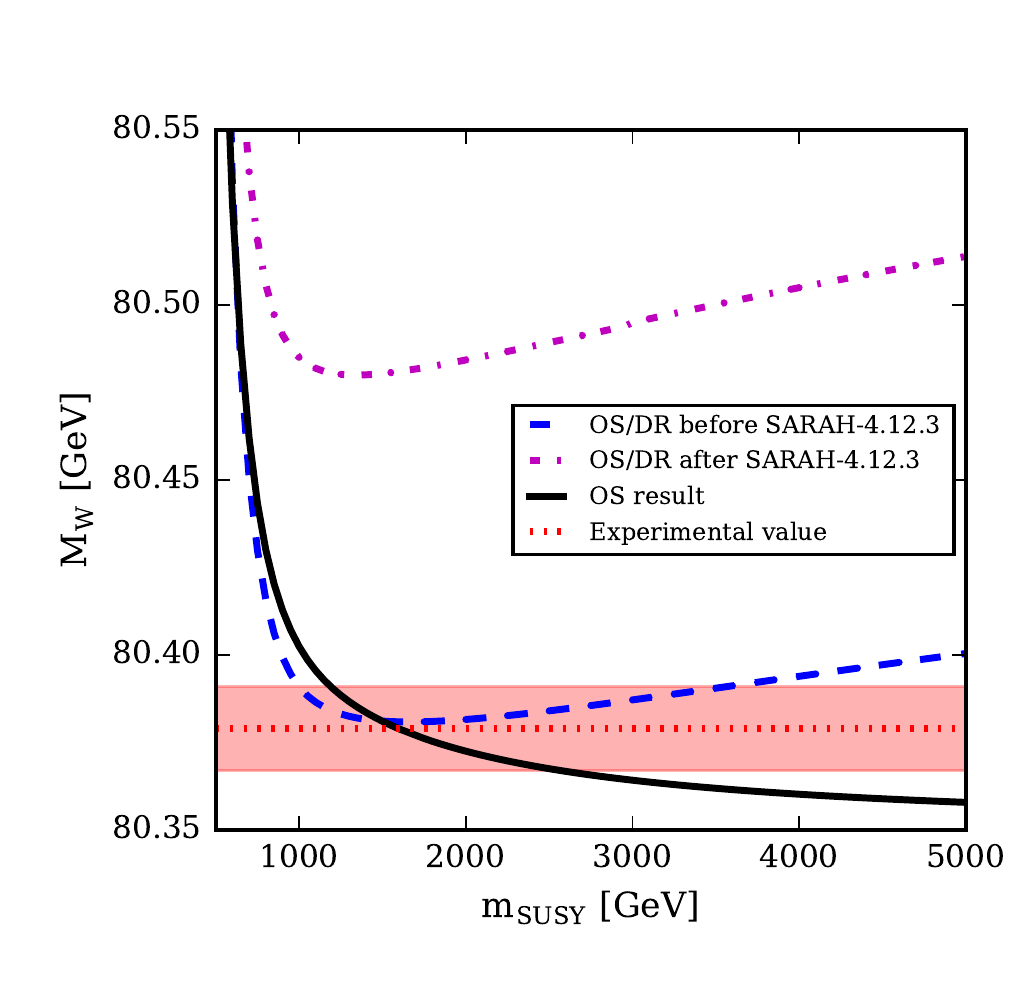}
\includegraphics[width=0.45\textwidth]{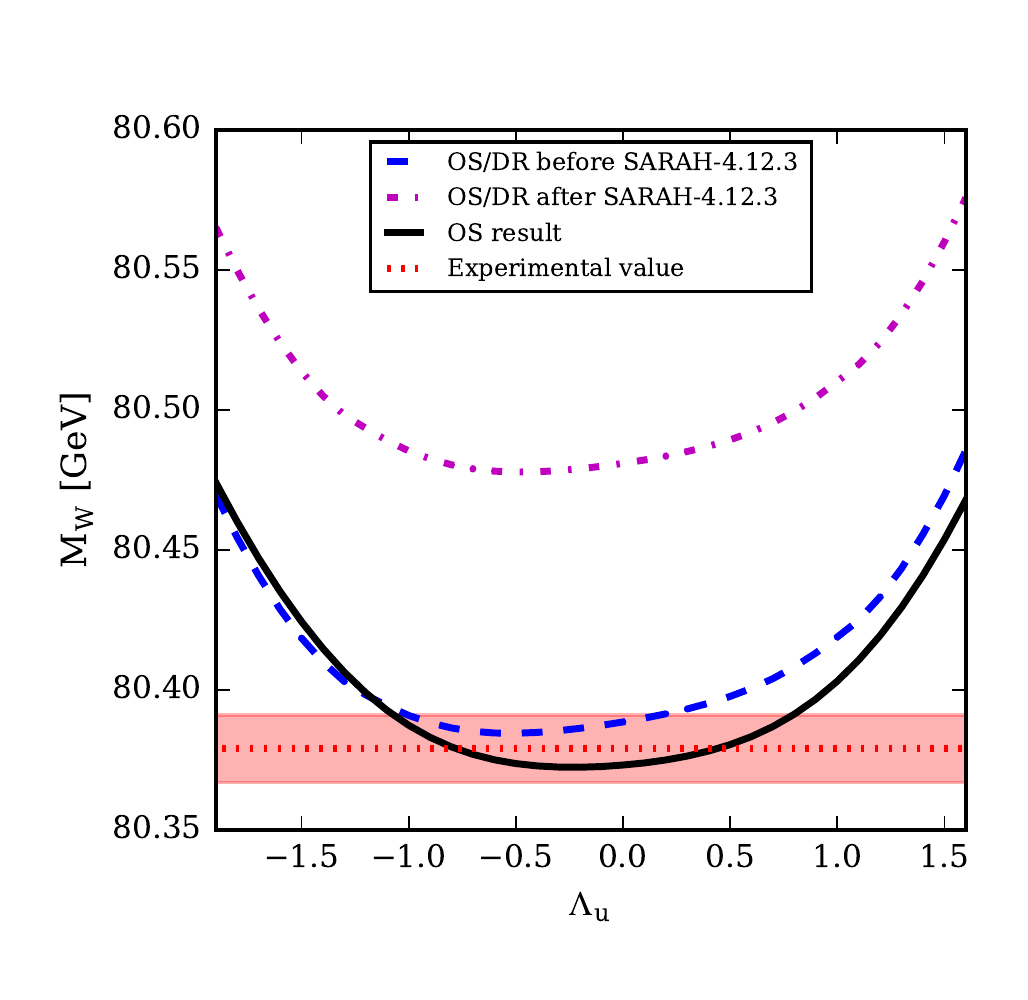}
\end{center}
\caption{Comparison of the $M_W$ prediction as a function of 
$m_{\text{SUSY}}$ (left) and $\Lambda_u$ (right) calculated in this work
using the 
on-shell scheme (black line) with the result from
\texttt{SARAH}/\texttt{SPheno} using 
the original contributions (blue dashed line) and 
using corrected contributions (purple dash-dotted line). 
As in the figures above, the red 
dotted line and the shaded area show 
the experimental central value for $\MW$
and its 1~$\sigma$ uncertainty band according to
eq.~\eqref{eq:mw_value_exp_all}.
All other model parameters are chosen as for BMP1 of ref.~\cite{PD2}.
For details see the discussion in the text.}
\label{fig:mw_lam_comparison}
\end{figure}

After the discussion of the various features of our numerical results, we
now turn to a comparison of our results 
with predictions for $\MW$ in the MRSSM that have
been obtained with other approaches.
As discussed above, the calculation of $M_W$ in refs.~\cite{PD1,PD2,PD3} 
was done using the implemented mixed OS/$\DR$ calculation of
\texttt{SARAH}/\texttt{SPheno}.
In ref.~\cite{Athron:2017fvs} and the update of the program \texttt{FlexibleSUSY}, 
it was recently shown that SM 
higher-order contributions implemented in \texttt{SARAH}/\texttt{SPheno}
were not correct with regard to the usage of the OS 
and $\DR$ top quark mass in the two-loop SM contribution.%
\footnote{
The tools \texttt{SPheno}, \texttt{SARAH}/\texttt{SPheno}
and \texttt{FlexibleSUSY} all contain the correct expressions in their latest versions
respectively.
} %
Correcting
the implementation leads to a shift of $M_W$ of 50--100 MeV compared 
to the result of the original calculation for example given in
refs.~\cite{PD1,PD2,PD3} for the MRSSM.
This is not only true for the MRSSM but also for all other versions of
\texttt{SARAH}/\texttt{SPheno} and also concerns the original 
\texttt{SPheno} code. In the following we show how the
on-shell calculation presented here compares to the
previous and corrected mixed OS/$\DR$ scheme predictions
obtained with \texttt{SARAH}/\texttt{SPheno}.

In figure~\ref{fig:mw_lam_comparison} we show the predictions for $M_W$
arising from the different calculations as function of the
common SUSY mass scale as defined in section~\ref{sec:gen_results} (left) 
and of the superpotential parameter $\Lambda_u$ of the MRSSM (right).
The on-shell result obtained in this work 
is given as solid black line, the previous OS/$\DR$ result
as dashed blue line and the corrected OS/$\DR$ result as dash-dotted 
purple line. The experimental result for $\MW$ is shown as red dotted line 
with a shaded region for the experimental 1~$\sigma$ uncertainty.

It can be seen in the left plot of figure~\ref{fig:mw_lam_comparison} 
that only the on-shell result presented in this work shows a proper
decoupling behaviour where for large $m_{\text{SUSY}}$ 
the MRSSM prediction approaches the SM prediction in which the Higgs boson
mass of the SM is identified with the mass of the SM-like Higgs boson of the
MRSSM, see also figure~\ref{fig:mw_msusy}. 
Both the previous and corrected mixed OS/$\DR$ scheme predictions
obtained with \texttt{SARAH}/\texttt{SPheno}
show a slope for large 
$m_{\text{SUSY}}$ such that the deviation between the MRSSM prediction and
the SM prediction grows for increasing $m_{\text{SUSY}}$. The 
correction of the contributions for
the OS/$\DR$ calculation yields a large 
upward shift of about 100~MeV in $\MW$ that is roughly constant for the
displayed range of $m_{\text{SUSY}}$. As a consequence, the prediction
for $\MW$ arising from the calculation in the
mixed OS/$\DR$ scheme including the correction shows a large deviation from
the measured value of $\MW$ for all values of $m_{\text{SUSY}}$.
The same feature also holds
for the plot as function of $\Lambda_u$ shown on the
right-hand side of figure~\ref{fig:mw_lam_comparison}. Also in this case the
correction that was implemented in the calculation based on the
mixed OS/$\DR$ scheme yields a roughly constant upward shift in $\MW$ that
brings the MRSSM prediction far away from the experimental result for $\MW$.
Both the previous and corrected \texttt{SARAH}/\texttt{SPheno} predictions
agree with the corresponding results from
\texttt{FlexibleSUSY} generator shown on the right of Fig. 10 in ref.~\cite{Athron:2017fvs}. 
Thus, the large deviation observed in the results for the MSSM and the MRSSM based on the mixed OS/$\DR$ scheme appears to be a general feature of this scheme.

In view of these findings the relatively good agreement between our on-shell 
result and the 
previous mixed OS/$\DR$ scheme prediction
obtained with \texttt{SARAH}/\texttt{SPheno} before the correction 
(dashed blue line) for $m_{\text{SUSY}}$ below $2$~TeV (left plot) and for 
the whole $\Lambda_u$ range (right plot) 
appears to be a numerical accident. This
accidental agreement implies that 
the parameter regions that were identified in 
refs.~\cite{PD1,PD2,PD3} for predicting $M_W$ values that are compatible
with the experimental result are roughly in agreement with the ones obtained
from our new on-shell prediction within the estimated uncertainties.
We have verified this accidental feature for the
benchmark points of the MRSSM proposed in refs.~\cite{PD1,PD2,PD3}.
Further investigations to clarify the observed features of the 
mixed OS/$\DR$ scheme prediction,
which has been first proposed for
the MSSM in ref.~\cite{BPMZ} and most generally been implemented in the tools \texttt{SARAH}/\texttt{SPheno} and \texttt{FlexibleSUSY}, 
would clearly be desirable.

\section{Conclusions}
\label{sec:conclusions}

We have presented the currently most accurate prediction
for the mass of the W boson in the MRSSM.
The result is based on the on-shell scheme,
using the muon decay 
constant, the fine-structure constant and the Z boson pole mass as precisely 
measured experimental inputs. The appearance of a triplet scalar 
vacuum expectation value $v_T$ 
at lowest order 
in the electroweak symmetry breaking condition 
for the W boson mass but not in the corresponding relation for the 
mass of the Z boson
leads to custodial symmetry breaking effects 
in the prediction for $M_W$ already at tree-level,
while the other BSM parameters of the MRSSM enter via higher-order corrections.
As described in detail, the prediction for the W boson mass 
needs to properly take into account 
the relation between the weak mixing angle and the gauge boson
masses that is modified in the MRSSM in comparison to the SM and extensions 
of it involving only Higgs doublets and singlets.

Our prediction for the W boson mass is based on the full 
one-loop contributions in the MRSSM that we have supplemented by all available 
SM-type corrections of higher order.
The implementation is based on a \texttt{SPheno}-4.0.3 mass spectrum generator 
that has been obtained by \texttt{SARAH}-4.12 for the MRSSM. 
The calculation of $M_W$ in the on-shell scheme has been carried out 
making use of the available analytical one-loop expressions that have been
consistently combined with the high-order corrections of SM-type. For the
latter a fit formula has been implemented, which accounts for the full 
electroweak two-loop corrections of SM-type involving numerical integrations
of two-loop integrals with non-vanishing external momenta, as well as
analytical results for leading QCD, electroweak and mixed corrections up to
the four-loop level.
For the renormalisation of the triplet scalar vacuum expectation 
value $v_T$ entering the prediction for $\MW$ 
at lowest order a 
$\overline{\text{DR}}$-type prescription has been chosen, where the
numerical value is determined via a two-step procedure ensuring numerical
stability.

We have investigated the numerical result for the W boson mass in
view of the characteristics of 
the parameter space of the MRSSM. 
%We have confirmed the most relevant parameter regions of the MRSSM
%identified in ref.~\cite{PD1} and extended the study of the effects.
We have verified in this context that in the decoupling limit where
the SUSY particles are heavy the SM prediction is recovered, i.e.\ in this
limit the MRSSM result approaches the SM prediction in which the Higgs boson
mass of the SM is identified with the mass of the SM-like Higgs boson of the
MRSSM. The upward shift in $\MW$ for a small SUSY mass scale tends to be more
pronounced in the MRSSM than it is the case in the MSSM and the NMSSM, while
the decoupling to the SM result occurs at higher values of the SUSY mass
scale than in those models.

The most relevant SUSY parameters of the MRSSM influencing the prediction
for $M_W$ are the triplet vacuum expectation value
and the $\Lambda$ superpotential couplings. 
The triplet vacuum expectation value is related to
all other model parameters through its contribution to the 
electroweak symmetry breaking conditions and 
gives rise to a lowest-order shift in the $\rho$ parameter.
Since the $\Lambda$ parameters are trilinear couplings of the superpotential,
%Therefore, it is expected that the prediction for $M_W$ depends 
%on them 
%This effect has been identified before~\cite{PD1} and we confirm it here.
they enter in a similar way as the Yukawa couplings, see also the 
discussion in ref.~\cite{PD1}. We have identified leading contributions to
the prediction for $\MW$ that enter with the fourth power of 
$\Lambda_u$ and $\Lambda_d$.
%The contributions to $M_W$ of the extended Higgs sector of the MRSSM to the
%ones of the squark sector we have seen a similar dependence both in shape
%and size. This can be explained by the common origin of the $\Lambda$ and
%the Yukawa couplings from the superpotential.
Additionally, the $\Lambda$ couplings have an important impact via the
contributions of the neutralino and chargino sector to $M_W$.
The contributions to $M_W$ of the extended Higgs sector of the MRSSM with a
common Higgs sector mass show a qualitatively similar behaviour as the ones
of the squark sector with a common squark mass.
Confronting the results of a parameter scan in the MRSSM as well as the SM
prediction for $\MW$ as a function of $m_t$ with the experimental results
for the W boson mass and the top quark mass, we have demonstrated a slight
preference for a non-zero SUSY contribution to $\MW$. While this preference
is similar to the results that were found in the MSSM and the NMSSM, it
should be noted that there is no direct limit from the MRSSM to the MSSM. 
This is caused in particular by the fact that
the $\Lambda$ couplings are a specific feature of the MRSSM and by
the pure Dirac nature of gauginos and Higgsinos in the MRSSM.
While certain contributions are similar in the
two models, in particular the MSSM-like contributions from stops and 
sbottoms which are driven by the top Yukawa coupling, the absence of
trilinear A-terms in the MRSSM also gives rise to differences in the squark
sector.

We have compared our results for $\MW$ to the ones that were obtained in the
MRSSM from
the mixed OS/$\DR$ implementation of \texttt{SARAH}/\texttt{SPheno}
before~\cite{PD1,PD2,PD3} and after the correction that was pointed out in 
ref.~\cite{Athron:2017fvs} was carried out. While as described above our
result shows a proper decoupling behaviour where for large values of a common SUSY
mass scale in the MRSSM the SM prediction is recovered, this is not the case
for either of the previous results. Those predictions show a deviation
from the SM result that actually grows with increasing $m_{\text{SUSY}}$.
While the result of the 
mixed OS/$\DR$ implementation of \texttt{SARAH}/\texttt{SPheno} 
{\em before the correction was applied\/}
agrees relatively well with our result in some parts of the parameter space,
a feature that is apparently caused by a numerical coincidence, 
the mixed OS/$\DR$ implementation of \texttt{SARAH}/\texttt{SPheno} 
{\em including the correction\/} 
shows a large upward shift in $\MW$ of about 100~MeV compared to the
previous result that gives rise to a large deviation from
the measured value of $\MW$.
Because of the described accidental numerical agreement our analysis 
roughly confirms the preferred MRSSM
parameter regions that were identified 
from the investigation of the $M_W$ prediction 
in refs.~\cite{PD1,PD2}. As we have demonstrated, the large deviations 
of the result of the  
mixed OS/$\DR$ implementation of \texttt{SARAH}/\texttt{SPheno} including
the correction both with respect to our result and with respect to the
experimental value of $\MW$ should motivate further work in this direction.

\section*{Acknowledgements}

We thank M.~Bach, W.~Kotlarski, W.~Porod, F.~Staub and D.~St\"ockinger for helpful discussions.
We acknowledge support
by the Deutsche Forschungsgemeinschaft (DFG, German Research Foundation)
under Germany's Excellence Strategy -- EXC 2121 ``Quantum Universe'' -- 
390833306.

\bibliographystyle{JHEP}
\bibliography{Literatur}

\end{document}